\documentclass[12pt]{article}

\AtBeginDocument{
  \addtocontents{toc}{\normalsize}
}

\pdfoutput=1

\usepackage{amsmath,amssymb,amsfonts,amsthm,amscd,mathrsfs}
\usepackage{xcolor}
\definecolor{darkblue}{rgb}{0.1,0.1,.7}
\usepackage[colorlinks, linkcolor=darkblue, citecolor=darkblue, urlcolor=darkblue, linktocpage]{hyperref} 
\usepackage[square, comma, compress,numbers]{natbib}
\usepackage[]{version}
\usepackage[]{graphicx}
\usepackage[]{latexsym}
\usepackage{geometry}
\usepackage[all,cmtip]{xy}
\usepackage[margin=10pt,font=small,labelfont=bf]{caption}
\usepackage{ifthen}
\usepackage{tikz}
\usepackage{array,setspace,mathrsfs,amsfonts,yfonts,dsfont,bbm,colonequals}
\usepackage{dsfont} 
\usepackage{xspace}
\usepackage{subcaption}

\usepackage{accents}
\newlength{\dhatheight}

\newcommand{\reef}[1]{(\ref{#1})}
\def\vareps{\varepsilon}
\def\eps{\epsilon}
\newcommand{\beq}{\begin{equation}} 
\newcommand{\eeq}{\end{equation}}
\def\del {\partial} 
\def\nn{\nonumber} 
\def\bZ {\mathbb{Z}} 
 
\def\bC {\mathbb{C}} 
\def\calO {{\cal O}}

\def\calN {{\cal N}} 
\def\calM {{\cal M}}

\def\calR {{\cal R}} 
\def\bZ {\mathbb{Z}} 
\def\bC {\mathbb{C}}

\def\ge{\geqslant}

\def\geq{\geqslant}
\def\leq{\leqslant}
\def\nn{\nonumber}
\def \del{\partial}
\def\Sd{\mathrm{S}_d}
\def\l{\ell} 
\def\eps{\epsilon}

\renewcommand{\d}{{\rm d}}

\newcommand{\be}{\begin{equation}}
\newcommand{\ee}{\end{equation}}
\def\ba{\begin{array}}
\def\ea{\end{array}}
\newcommand{\mysigma}{s}
\newcommand{\vev}[1]{\ensuremath{\langle #1 \rangle}\xspace}
\newcommand{\op}{\ensuremath{\mathcal{O}}\xspace}
\newcommand{\G}{\Gamma}
\newcommand{\D}{\Delta}

\newcommand\restr[2]{{\left.\kern-\nulldelimiterspace#1\vphantom{\big|}\right|_{#2}}}

\newcommand\zb{\bar{z}}

\newcommand{\bea}{\begin{eqnarray}}
\newcommand{\eea}{\end{eqnarray}}

\numberwithin{equation}{section}
\begin{document}

\vspace*{-.6in} \thispagestyle{empty}
\begin{flushright}
CERN PH-TH/2017-052\\
YITP-SB-17-13
\end{flushright}
\vspace{1cm} {\large
\begin{center}
{\bf A scaling theory for the long-range to short-range crossover\\
and an infrared duality}\\
\end{center}}
\vspace{1cm}
\begin{center}
{\bf Connor Behan$^{a}$, Leonardo Rastelli$^{a}$, Slava Rychkov$^{b,c}$, Bernardo Zan$^{d,b}$}\\[2cm] 
{
\small
$^{a}$ C.N. Yang Institute for Theoretical Physics, Stony Brook University, \\ Stony Brook, NY 11794, USA
\\
$^b$  CERN, Theoretical Physics Department, 1211 Geneva 23, Switzerland\\
$^c$  Laboratoire de Physique Th\'eorique de l'\'Ecole Normale Sup\'erieure,  \\
PSL Research University,  CNRS,  Sorbonne Universit\'es, UPMC
Univ.\,Paris 06,  \\
24 rue Lhomond, 75231 Paris Cedex 05, France
\\
$^d$ Institut de Th\'eorie des Ph\'enom\`enes Physiques, EPFL, CH-1015 Lausanne, Switzerland\\
\normalsize
}
\end{center}

\vspace{4mm}
\begin{center}
{\it To John Cardy on his 70th birthday.}
\end{center}
\begin{abstract}
We study the second-order phase transition in the $d$-dimensional Ising model with long-range interactions decreasing as a power of the distance $1/r^{d+s}$. For $s$ below some known value $s_*$, the transition is described by a conformal field theory without a local stress tensor operator, with critical exponents varying continuously as functions of $s$. At $s=s_*$, the phase transition crosses over to the short-range universality class. While the location $s_*$ of this crossover has been known for 40 years, its physics  has not been fully understood, the main difficulty being that the standard description of the long-range critical point is strongly coupled at the crossover. In this paper we propose another field-theoretic description which, on the contrary, is weakly coupled near the crossover. We use this description to clarify the nature of the crossover and make predictions about the critical exponents. That the same long-range critical point can be reached from two different UV descriptions provides a new example of infrared duality.
\end{abstract}

\hspace{0.7cm} March 2017

\newpage

{
\setlength{\parskip}{0.04in}
\renewcommand{\baselinestretch}{0.4}\normalsize
\tableofcontents
\renewcommand{\baselinestretch}{1.0}\normalsize
}


\setlength{\parskip}{0.1in}
\newpage

\section{Introduction} 

Spin models with long-range interactions exhibit rich critical behavior with continuously varying exponents. 
In this paper we  focus on the ferromagnetic long-range Ising model (LRI), with spin-spin interaction decaying as a power\footnote{The exponent $s$ is usually denoted $\sigma$, but here we reserve letter $\sigma$ for the short-range Ising spin field.}  of the distance $\sim 1/r^{d+s}$. The lattice Hamiltonian is given by
\beq
\label{eq:LRI}
H_s = - \sum_{i,j} \frac{J}{|i-j|^{d+s}}S_i S_j,
\eeq
where $S_i=\pm 1$ are the Ising spin variables, and $J>0$ in the considered ferromagnetic case. 
We will assume $s>0$ for the thermodynamic limit to be well defined. The space dimensionality $d$ will be $d=2,3$. Formally our considerations will apply also to non-integer dimensions in the range $1<d<4$.

Forty years of theoretical considerations and Monte Carlo simulations have established that model \reef{eq:LRI} has a second-order phase transition for each $s>0$, whose nature however varies with $s$. One distinguishes three critical regimes: (i) the mean-field regime for $s<d/2$, (ii) the intermediate regime for $s>d/2$ and up to a certain $s_*$ (see below) and (iii) the short-range regime $s>s_*$. {The primary goal of this paper will be to elucidate the long-range to short-range crossover at $s=s_*$.}\footnote{\label{note:d=1}This also explains why $d=1$ is excluded from consideration: for $d=1$ we have $s_*=1$, but for $s>s_*$ phase transition is absent. We will comment on 
the $d=1$ case in the discussion section.} A short summary of our results has appeared in \cite{short}.

Let us start by briefly reviewing the regimes (i)-(iii). To study the long-distance behavior, it's standard to replace the lattice model with a continuum field theory in the same universality class. 
Besides the usual quadratic and quartic local terms, the action includes a gaussian non-local term (with a negative sign for the ferromagnetic 
interaction)\footnote{\label{note:0s2}We will assume $0<s<2$, which includes the long-range to short-range crossover point $s_*$, see \reef{s*}. This action is appropriate in this interval. Beyond $s=2$ the local kinetic term $(\partial \phi )^2$ becomes relevant and would have to be added.}
 \beq
 S= -\int d^dx\, d^dy\, \frac{\phi(x)\phi(y)}{|x-y|^{d+s}} + \int d^d x [t\,\phi(x)^2 + g\, \phi(x)^4]\,.
 \label{standardflow}
 \eeq
The non-local term by itself describes mean field theory (MFT); it endows $\phi$ with dimension\footnote{Scaling dimensions of various fields $X$ are denoted interchangeably by $[X]$ or $\Delta_X$.} 
\beq
[\phi]_{\rm UV}=(d-s)/2.
\eeq 
The quadratic term is always relevant and its coefficient $t$ must be tuned to zero to reach the transition.\footnote{{\label{note:dis}See appendix \ref{sec:dis} for a discussion of physics of the long-range Ising model away from the transition.}
} The quartic term is irrelevant for $s<d/2$, explaining why the transition is mean-field in this range. For $s>d/2$ the quartic induces a nontrivial renormalization group (RG) flow. In the intermediate regime $d/2<s<s_*$, the flow ends in an interacting long-range fixed point (LRFP). The composite operators such as $\phi^2$, $\phi^4$ acquire nontrivial anomalous dimensions, as befits an interacting fixed point. However, the dimension of $\phi$ is controlled by a non-local term and does not get renormalized at the fixed point: 
\beq
[\phi]_{\rm LRFP}=[\phi]_{\rm UV}=(d-s)/2\,.
\eeq 
Finally, the crossover from the long-range to the short-range regime happens \cite{Sak} when $[\phi]_{\rm LRFP}$, decreasing with $s$, reaches the short-range Ising fixed point (SRFP) dimension $[\phi]_{\rm SRFP}$. In other words, the dimension of $[\phi]$ varies continuously through the crossover. This fixes 
\beq
s_*=d-2[\phi]_{\rm SRFP} = 2-\eta
\label{s*}
\eeq 
in terms of the SRFP critical exponent $\eta$. Although we use the word ``crossover'', it's important to emphasize that the transition happens sharply, at $s=s_*$.

\begin{figure}[h]
\centering
\includegraphics[width=0.5\textwidth]{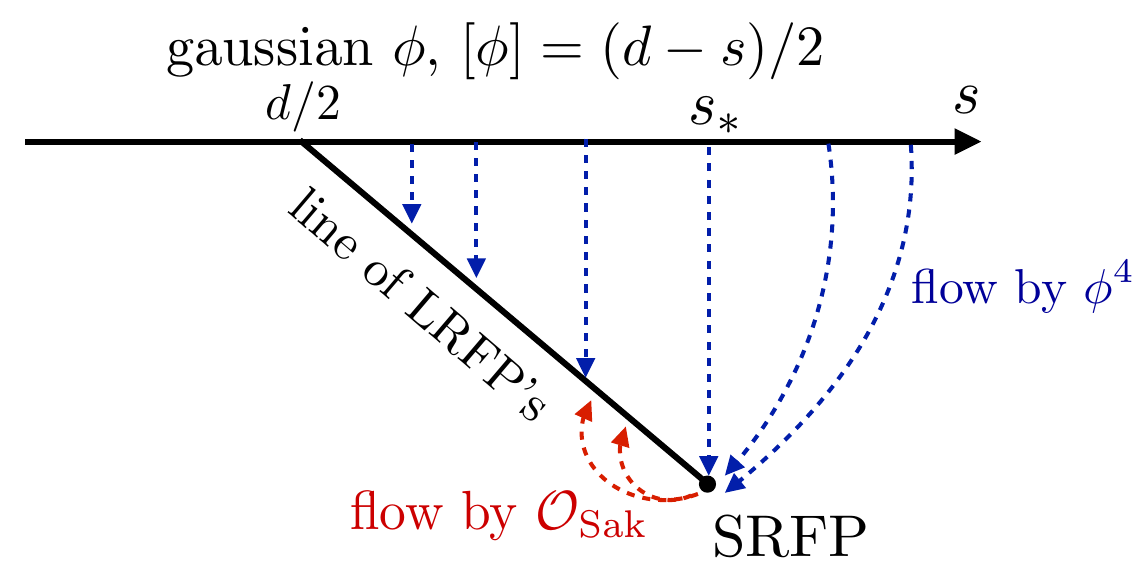}
\caption{RG flow diagram of the standard picture.}
\label{fig-standard}
\end{figure}

The physical picture that we just reviewed is commonly accepted since the original work by Fisher et al. \cite{Fisher:1972zz} and its refinement by Sak 
\cite{Sak,Sak77}. We'll refer to it as ``standard".
However, while the crossover from the mean-field to the intermediate regime is well understood, some features of the long-range to short-range crossover remain puzzling. For $s$ slightly above $d/2$, the quartic interaction is slightly relevant and one can study the flow perturbatively, computing physical quantities in a systematic expansion in $\eps = 2s-d$. By contrast, a perturbative description of the long-range to short-range crossover is presently lacking. Sak \cite{Sak77} (see also Cardy's book \cite{Cardy:1996xt}, section 4.3),
proposed to analyze the SRFP stability in terms of the non-local perturbation 
\beq 
\label{eq:cardy}
\calO_{\rm Sak}=\int d^dx\, d^dy\, \frac{\sigma(x)\sigma(y)}{|x-y|^{d+s}}\,,
\eeq 
where $\sigma\equiv \phi_{\rm SRFP}$ is the SRFP spin field. This perturbation crosses from relevant to irrelevant precisely at $s=s_*$ \cite{Sak77,Cardy:1996xt}. For $s<s_*$ the SRFP perturbed by $\calO_{\rm Sak}$ should flow to the LRFP. The RG flow diagram summarizing the standard picture is shown in Fig.~\ref{fig-standard}. If $s$ is just slightly below $s_*$, Sak's perturbation is weakly relevant, and in principle it should be possible to study the flow perturbatively. However, it is unclear how to adapt the rules of conformal perturbation theory to this non-local case.
To the best of our knowledge this has not been done. 

This lack of computability may be dismissed as a technical problem, but there are related conceptual puzzles. If the crossover is continuous, the spectrum of all operators, not just $\phi$, should vary continuously.
In particular, the number of operators should be the same on both sides of the crossover. However, for some LRFP operators no counterpart SRFP operators appear to exist. 

One such operator is $\phi^3$. It has been shown in \cite{Paulos:2015jfa} (we will review the argument below) that the dimension of this operator at the LRFP satisfies the ``shadow relation":
\beq
\label{sec:shadow0}
[\phi^3]_{\rm LRFP}+[\phi]_{\rm LRFP}=d\,.
\eeq
This suggests that at the crossover point there should be a $\bZ_2$ odd operator of dimension $d-[\phi]_{\rm SRFP}$. This is puzzling, because the SRFP Ising contains, both in $d=2$ and $d=3$, a \emph{single} relevant $\bZ_2$ odd scalar.

Another puzzle involves the stress tensor operator. The SRFP has a local conserved stress tensor $T_{\mu\nu}$. Moving to the long-range regime, this operator is expected to acquire an anomalous dimension so that it's no longer conserved. The divergence $V_{\nu}=\del^\mu T_{\mu\nu}$ is thus a nontrivial operator at the LRFP. At the crossover point the dimension of this vector operator is exactly $d+1$. Is there such an operator in SRFP? 
For $d=2$ the SRFP is a solvable minimal model conformal field theory (CFT), and it's easy to see by inspection that there is no such operator. Fir $d$ close to 4 one can use the weakly coupled Wilson-Fisher description, and again there is no such operator. While in $d=3$ its existence cannot be rigorously excluded at present,\footnote{
$\bZ_2$-even operators of odd spin have not yet been probed by the numerical conformal bootstrap.} it seems very unlikely. 

The puzzle of the missing $V_\mu$ can be stated more formally in terms of ``recombination rules'' of unitary representations of the ${\mathfrak s\mathfrak o}(d+1, 1)$ conformal algebra.\footnote{It is common lore that SRFP is conformally invariant, in any $d$. The fact that scale invariance enhances to $SO(d+1,1)$ conformal invariance at the LRFP -- even in the absence of a local stress tensor -- has been recently demonstrated in \cite{Paulos:2015jfa}, using the flow \reef{standardflow}. We will review this result in section \ref{sec:standard}.} Now, the standard  stress tensor of the SRFP is the lowest weight state (conformal primary) of the {\it shortened} spin-two representation ${\cal C}_{\ell = 2}^{d}$, while the non-conserved spin-two operator of the LRFP is the conformal primary of the {\it long} spin-two representation ${\cal A}_{\ell = 2}^\Delta$, with $\Delta \geq d$ for unitarity.
When the unitarity bound is saturated, the long spin-two representation decomposes into the semi-direct sum
${\cal A}_{\ell = 2}^d  \simeq {\cal C}_{\ell = 2}^{d} \oplus {\cal A}_{\ell = 1}^{d+1}$.  In other terms, the shortened spin-two representation can become long only by recombining
with (``eating") an additional spin-one representation,  ${\cal A}_{\ell = 1}^{d+1}$, whose conformal primary
 $V_\mu$ is however missing in the SRFP.

In this paper we will propose a modified theory of the long-range to short-range crossover, which will both resolve the puzzle of missing states and lead to concrete predictions of how the long-range exponents vary near the crossover point. 

\subsection{Our picture}

The need to resolve the above-mentioned difficulties leads us to the following modified picture of the LRFP to SRFP crossover (referred to as ``our picture" below). Like in the standard picture, the crossover in our picture does happen continuously and at $s=s_*$. However, and this is where we differ, we posit that LRFP crosses over not to SRFP, but to a larger theory, which consists from SRFP and a decoupled sector: the mean-field theory of a gaussian field $\chi$. This larger theory will be referred to as ``SRFP+$\chi$". The two point (2pt) function of $\chi$ will be taken unit-normalized: $1/|x|^{2\Delta_\chi}$. 

Assuming our picture, we can construct the flow from the ``SRFP+$\chi$'' theory to the LRFP by turning on the perturbation
\beq
\label{eq:ourpert}
g_0 \int d^dx\, \calO(x),\quad \calO = \sigma\cdot  \chi\,.
\eeq
The sign of $g_0$ is arbitrary since it can be flipped by the $\bZ_2^\chi$ symmetry $\chi\to-\chi$. In fact the decoupled SRFP+$\chi$ theory has an enlarged $\bZ_2^\sigma\times \bZ_2^\chi$ symmetry which is broken to the diagonal when the perturbation $\calO$ is turned on. This is as it should be, since the LRFP has only a single $\bZ_2$ symmetry $\phi\to-\phi$. The enlarged symmetry of SRFP+$\chi$ leads to selection rules, which will appear many times in the RG calculations below.

Connection to the standard picture is established by integrating out $\chi$, which should generate precisely Sak's non-local perturbation \reef{eq:cardy}.\footnote{Notice that for real $g_0$, the generated $\calO_{\rm Sak}$ has ferromagnetic, negative, sign, as it should.} This fixes the dimension $[\chi]=(d+s)/2$, so that
\beq
[\calO]=[\chi]+[\sigma]=d-\delta,\quad \delta = (s_*-s)/2\,.
\eeq 
This crosses from relevant to irrelevant at the same location as before.  We emphasize however that $\chi$ is not simply a theoretical construct introduced to represent $\calO_{\rm Sak}$, but is a physical field. 

One way to think about $\chi$ is that it's a remnant of the long-range interactions present in the original model \reef{standardflow}. {Notice also that $[\chi]$ coincides with $[\phi]_{\rm dis}$, the dimension of $\phi$ in the disordered phase of the long-range model; see appendix \ref{sec:dis}.} This suggests a similarity between the disordering effects of temperature and of the short-range critical fluctuations on the long-range correlations.

It should be possible to verify the existence of $\chi$ via lattice measurements. This is true even in the short-range regime $s>s_*$, where it is decoupled. The point is that it is decoupled from the SRFP scaling fields, but not from the lattice operators. It should thus be possible to detect $\chi$ by measuring the spin-spin correlation function $\langle S_i S_j\rangle$ on the lattice. At the critical point and at large distances, this function has a power law expansion of the form
\beq
\langle S_i S_j\rangle \sim \sum c_k/r^{2\Delta_k}\,,\quad r=|i-j|\,,
\eeq
where $\Delta_k$ are the dimensions of $\bZ_2$-odd scalar operators. We predict that at the LRFP the dimension of $\chi$ should appear among $\Delta_k$.

The existence of $\chi$ allows to resolve the difficulties concerning the crossover description. First of all, since $\chi$ and $\phi$ satisfy the shadow relation $[\chi]+[\phi]=d$, $\chi$ can be identified with $\phi^3$ at the crossover point. This identification and its consequences will be discussed in detail below. We will also see that using $\chi$ one can construct a vector operator playing the role of $V_\mu$. Finally, since $\calO$ is a local operator, we will be able to use the well-developed framework of conformal perturbation theory to compute the long-range critical exponents near the crossover point. 

\begin{figure}[t]
\centering
\includegraphics[width=0.5\textwidth]{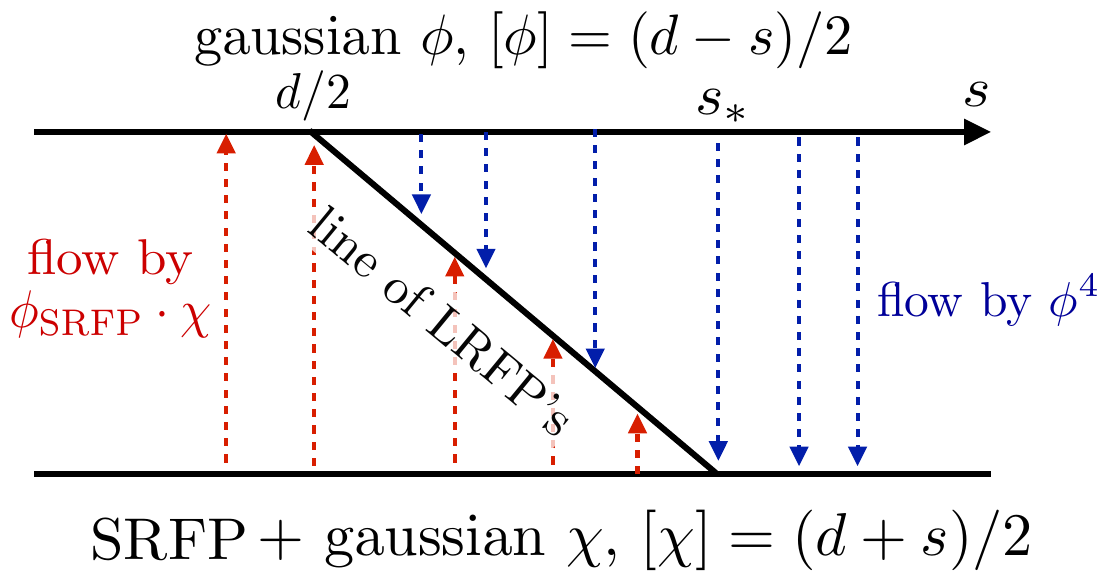}
\caption{RG flow diagram of our picture.}
\label{fig-ours}
\end{figure}

The RG flow diagram of our picture is shown in Fig.~\ref{fig-ours}. We predict that in the intermediate regime $d/2 < \mysigma < \mysigma_*$ the LRFP is the common IR endpoint of two distinct RG flows: 
\begin{enumerate}
\item Flow (\ref{standardflow}) from the mean field theory, which is weakly coupled near the lower end of the intermediate regime $(\epsilon \to 0$). We will call this ``$\phi^4$-flow".

\item Our newly proposed flow emanating from the SRFP+$\chi$ theory, which is weakly coupled near the crossover $(\delta \to 0$). We will call this ``$\sigma\chi$-flow".
\end{enumerate}

In quantum-field theoretic parlance, this situation -- when the same IR theory can be reached from two different UV descriptions -- is referred to as ``infrared duality".
A famous example is the Seiberg duality which establishes the IR equivalence of UV-distinct $\calN=1$ supersymmetric gauge theories \cite{Seiberg:1994pq}. Another example is the particle/vortex duality between the $XY$ model and the $U(1)$ Abelian Higgs model in 3d, both flowing to the same $O(2)$ Wilson-Fisher critical point \cite{Peskin,PhysRevLett.47.1556}. The novelty of our example is that the IR fixed point does not have a local stress tensor.

\subsection{Outline}

We will start by investigating the structure of the LRFP close to the crossover ($\delta\ll 1$). In this region our $\sigma\chi$-flow provides a weakly coupled description of the LRFP, allowing a number of explicit computations. In section 
\ref{sec:beta} we compute the beta-function, at the leading nontrivial order, and establish the fixed point existence. In section \ref{sec:anom} we compute the leading anomalous dimensions for the most interesting operators. In particular, we demonstrate that the stress tensor acquires anomalous dimension, and exhibit the eaten operator $V_\mu$, thus resolving the paradox of missing states. These two sections include also a self-consistent presentation of necessary conformal perturbation theory techniques.

In section \ref{sec:standard} we review what is known about the $\phi^4$-flow, mostly following Ref.~\cite{Paulos:2015jfa}. We discuss a number of results valid to all orders in perturbation theory (the absence of the anomalous dimension of $\phi$, the shadow relation between $\phi$ and $\phi^3$). We also give a streamlined exposition of the proof of the conformal invariance of the LRFP.
Finally, we discuss the relation between the OPE coefficients of $\phi$ and $\phi^3$ which can be proved using the non-local equation of motion (EOM).

We start section \ref{sec:IRduality} with some all-order results for the $\sigma\chi$-flow, established by analogy with the $\phi^4$-flow. We then show that the results about the LRFP obtained from the $\phi^4$-flow and the $\sigma\chi$-flow match together beautifully, providing compelling evidence for the infrared duality. At this high point, we conclude.

We also have a number of appendices. Appendix \ref{sec:dis} is devoted to the discussion of the long-range Ising model off criticality. Appendix \ref{sec:prior} is an (admittedly incomplete) review of our favorite works on the long-range Ising model, both in the physics and the mathematics literature. Appendix \ref{sec:integral} evaluates some integrals arising in the conformal perturbation theory calculations.
Appendix \ref{sec:OPE} checks explicitly a curious conclusion, reached in Ref.~\cite{Paulos:2015jfa} and reviewed in section \ref{sec:confinv}, that scale invariance of the IR fixed point of the $\phi^4$ flow implies that the kernel of the $\phi\times\phi^4$ OPE should integrate to zero. Finally, 
Appendix \ref{sec:checksOPE} provides some sanity checks of the relations between the OPE coefficients which follow from the non-local EOM, both for the $\phi^4$ and for the $\sigma\chi$-flow.

\section{Beta-function}
\label{sec:beta}

According to our proposal, the LRFP can be described as the IR fixed point of the $\sigma\chi$-flow. This description is weakly coupled for $\delta\ll 1$, when the $\sigma\chi$ perturbation is weakly relevant. This allows to compute the LRFP critical exponents in terms of the SRFP conformal data, known exactly in $d=2$ \cite{Belavin:1984vu}, and with an impressive precision in $d=3$ thanks to the recent progress in the numerical conformal bootstrap. 

The standard framework to describe CFTs with turned on weakly relevant local perturbations is conformal perturbation theory (see e.g.~\cite{Zamolodchikov:1987ti,Cardy:2008jc} for $d=2$ and \cite{Cardy:1996xt,Cappelli:1991ke} for general $d$). 
As usual in quantum field theory, we consider the perturbative expansion of observables in the bare coupling constant in a regulated theory, and then add counterterms to cancel the dependence on the short-distance regulator. The order $n$ perturbative correction to an observable $\Xi$ is given by
\begin{equation}
\label{Ogn}
\frac{g_0^n}{n!}
\int d^dx_1\ldots d^dx_n 
\langle \calO(x_1)\ldots\calO(x_n)\,\Xi\rangle\,.
\end{equation}
In general, this integral is divergent when points $x_i$ collide. A convenient way to regulate is by point splitting, restricting integration to the region where all $|x_i-x_j|>a$ (short-distance cutoff). If $\Xi$ is a local operator, there will also be divergences where $x_i$ approach $\Xi$, but those are associated not with the running of the coupling but with the renormalization of $\Xi$. They will be discussed and interpreted separately below. 

The first quantity we need is the beta-function. Let $g= a^{\delta} g_0$ be the dimensionless coupling at the cutoff scale. The beta-function has the form
\beq
\beta(g)\equiv \frac{d g}{d\log (1/a)}=-\delta\, g +\ldots\,,
\eeq
where $-\delta\, g$ is the classical term and $\ldots$ are the quantum corrections. 

The order $g^2$ correction to the beta-function is proportional to the 3pt function coefficient $C_{\calO \calO \calO}$. This is well known and sufficient for most applications \cite{Zamolodchikov:1987ti,Cardy:2008jc,Cardy:1996xt}. However, in our case $C_{\calO\calO\calO}$ vanishes, because $\calO$ is odd under the $\bZ_2^\chi$ symmetry $\chi\to-\chi$. Analogously all even-order contributions to $\beta(g)$ will vanish as well. 

The lowest nonvanishing contribution will appear at order $g^3$, at that's the only one we will use in this work. So we will have:
\beq
\beta(g) = -\delta g+\beta_3 g^3\,,
\eeq
neglecting the higher order terms. We will now review how one computes the coefficient $\beta_3$. We aim to discuss the fixed point properties at the leading nontrivial order in $\delta$. For this we may neglect the dependence of $\beta_3$ on $\delta$, so we will compute it in the limit $\delta=0$.\footnote{To compute higher order corrections, we would have to keep $\delta$ nonzero and set up a minimal subtraction scheme. This will
not be carried out in this work, although see the all-order discussion in section \ref{sec:all-order}.} We will also specialize to the case $C_{\calO \calO \calO}=0$ of interest to us, as this simplifies some details. See \cite{Gaberdiel:2008fn,Poghossian:2013fda} for prior work involving third-order corrections. Our discussion owes a lot to \cite{Komargodski:2016auf}, which covers also the general case $C_{\calO \calO \calO}\ne0$. 
 
For $\delta=0$ the coupling $g$ is marginal and its running is related to the logarithmic short-distance divergence of \reef{Ogn}. At order $g^3$, we are interested in the divergence where three points come close together. In this region we can use the `triple operator product expansion (OPE)': 
\beq
\label{eq:teripleOPE}
\calO(0)\calO(x_2)\calO(x_3)\sim f(x_2,x_3) \calO(0)\,.
\eeq
It's easy to see that $f(x_2,x_3)$ is nothing but the 4pt correlation function:\footnote{As usual $\calO(\infty)=\lim_{x\to\infty} |x|^{2\Delta_\calO}\calO(x).$} 
\beq
\label{eq:f}
f(x_2,x_3) = \langle \calO(0) \calO(x_2) \calO(x_3)\calO(\infty)\rangle\,.
\eeq
This is similar to the well-known relation between the usual OPE of two operators and the 3pt function. To check \reef{eq:f}, use \reef{eq:teripleOPE} in the r.h.s.~and the fact that $\langle \calO(0) \calO(\infty)\rangle=1$.

Using \reef{eq:teripleOPE}, we see that the divergence of the integral with three $\calO$ insertions is equal to the integral with one $\calO$ insertion times a divergent coefficient, computed by integrating the 4pt function:
\beq
\label{eq:intf}
\int_V d^d x_1\, d^d x_2\,d^d x_3\,\langle \calO(x_1) \calO(x_2) \calO(x_3)\calO(\infty)\rangle = A V \log (1/a)+\ldots 
\eeq
Integration is over the region $|x_i-x_j|>a$ with all three points belonging to a finite region of volume $V$, which serves as an IR cutoff. The IR cutoff is needed since we are interested only in the short-distance part of the divergence.  

The divergence at $O(g^3)$ can thus be canceled, and the cutoff dependence removed, by a variation of the $O(g)$ term, adjusting the bare coupling by $-A \log (1/a)\times (g^3/3!)$. Therefore, the beta-function is given, to this order, by $\beta(g)=\beta_3 g^3$ with 
\beq
\label{eq:b3}
\boxed{\beta_3= - A/3!}\,.
\eeq  

To isolate the coefficient $A$, we use translational invariance to fix one of the points, say $x_3$, to $0$. The volume factor $V$ cancels, and we are left with an integral of the function $f(x_1,x_2)$. We then separate the integration over the overall `size' of the pair of points $(x_1,x_2)$ and over their relative position. Rescaling the pair by, say, $|x_1|$, and using the fact that $f$ has dimension $2d$ we have
\begin{align}
\int d^d x_1\,d^d x_2 \, f(x_1,x_2) = \int d^d x_1\,d^d x_2\frac 1{|x_1|^{2d}} f\left (\frac{x_1}{|x_1|},\frac{x_2}{|x_1|}\right)={\rm S}_{d} \int \frac{d|x_1|}{|x_1|}\int d^d y\, f (\hat {e},y)\,,\label{eq:AS1}
\end{align}
where $\hat{e}$ is an arbitrary unit length vector and ${\rm S}_{d}=2\pi^{d/2}/\Gamma(d/2)$ is the volume of the unit sphere in $d$ dimensions. The log divergence $\sim \log(L/a)$ now arises from integrating over $a<|x_1|<L$, which is basically the pair size. So we conclude
\beq
\label{eq:Afinal}
A = {\rm S}_{d} \int d^d y\, f (\hat {e},y)\,\qquad\text{(naive)}.
\eeq
As written this expression is naive, since the $y$ integral must be defined with care. In principle, the $y$ integral in \reef{eq:AS1} was meant to be computed with a UV cutoff $a/|x_1|$. If the $y$ integral were convergent, we could simply extend the integration to the whole space, as this does not affect the logarithmic divergence that we are after. However, the integral is not in general convergent, and this complicates matters.

The complication can be traced to the fact that, in the above discussion, we neglected that the integral \reef{eq:intf} contains power divergences on top of the log divergence.
These power divergences have nothing to do with the running of $g$. Instead, they renormalize coefficient of the relevant operators appearing in the OPE $\calO\times\calO$. In our case there are two such operators, the unit operator and the SRFP energy density operator $\vareps$.\footnote{Another low-dimension scalar operator in the $\calO\times\calO$ OPE is $\chi^2$, but this one is irrelevant since $s>0$.} The unit operator coefficient is unimportant, while that of $\varepsilon$ has to be anyway tuned to zero to reach the fixed point, as this corresponds to tuning the temperature to the critical temperature. The bottom line is that the power divergences need to be subtracted away. 

There are two methods to do this, which give equivalent, although not manifestly identical, final results. 
Method 1 subtracts the divergent terms, given by the relevant operators, from the integrand $f$. Method 2 computes the integral \reef{eq:Afinal} with a cutoff and drop the terms that diverge when the cutoff is sent to zero. In both case we are just dropping power divergences of the integral \reef{eq:AS1}, and we are not changing the coefficient of the logarithm divergence. Once one of these methods has been employed, the integral is convergent.

\emph{Method 1.} We subtract from the integrand $f$ in \reef{eq:AS1} 
the singularities associated with the two relevant operators in the limits $x_1\to0$, $x_2\to0$, $x_1\to x_2$. The subtraction terms have to be chosen so that they fully subtract the power divergence but do not modify the logarithmic divergence. The following simple choice satisfies these constraints:
\begin{gather}
\label{eq:subterms}
f\to \tilde f = f-r_1-r_\vareps\,,\\
r_1 = \frac{1}{|x_1|^{2d}}+ \frac{1}{|x_2|^{2d}}+\frac{1}{|x_1-x_2|^{2d}}\,,\nn\\
r_\vareps = (C_{\sigma\sigma\vareps})^2\left(\frac{1}{|x_1|^{2d-\Delta_\vareps}|x_2|^{\Delta_\vareps}}
+\frac 1{|x_2|^{2d-\Delta_\vareps}|x_1|^{\Delta_\vareps}}
+\frac 1{|x_1-x_2|^{\Delta_\vareps}|x_1|^{2d-\Delta_\vareps}}\right)\,.\nn
\end{gather}
Here $C_{\sigma\sigma\vareps}$ is the SRFP OPE coefficient: $\sigma\times\sigma = \mathds{1}+C_{\sigma\sigma\vareps}\vareps+\ldots.$ The crucial point is that these subtraction terms themselves only have power divergences. This is obvious for $r_1$. For $r_\vareps$, notice that the $d^dx_1 d^d x_2$ integral of each term factorizes into a product of two integrals each of which has only power divergences. So the logarithmic divergence is not modified by the subtraction procedure.

The regulated expression for $A$ is then obtained by $f\to \tilde f$ in \reef{eq:Afinal}:\beq
\label{eq:Afinalt}
\boxed{A = {\rm S}_{d} \int d^d y\, \tilde f (\hat {e},y)}\,.
\eeq
This integral is now convergent, although not absolutely convergent. The lack of absolute convergence is due to the presence of relevant or marginal operators with nonzero spin in the $\calO\times\calO$ OPE. These are $\del_\mu\vareps$ and the stress tensor $T_{\mu\nu}$. Since these operators have nonzero spin, their contributions vanish when integrated over the angular directions. So the integral has to be understood in the sense of principal value, introducing and then removing spherical cutoffs around 0, $\hat e$ and $\infty$. These cutoffs are remnants of the original cutoffs on $|x_2|$ and $|x_1-x_2|$, since $y$ is the rescaled $x_2$.

\emph{Method 2.} In this method we start by spliting the integration region of \reef{eq:intf}
 into three parts. We consider one region in which $x_{12}$ is the shortest distance:
\beq
\mathcal{R}_{12}=\left\lbrace x_1,x_2,x_3 : |x_{12}|<|x_{13}|, |x_{12}|<|x_{23}| \right\rbrace \,,
\label{eq:regionR}
\eeq
and the two other regions $\mathcal{R}_{23}$ and $\mathcal{R}_{13}$, given by permutations of the three points. It is clear that these three regions contribute equally to the integral \reef{eq:intf}, so we can focus on $\mathcal{R}_{12}$.
As before, we set one of the points to zero and we rescale $x_1$ and $x_2$ by $|x_1|$. The logarithmic divergence arises when integrating over $|x_1|$. We obtain
\beq
\boxed{A=3 {\rm S}_{d}\int_\mathcal{R} d^d y\, f (\hat {e},y)}\,,
\label{eq:A2method}
\eeq
where
\beq
\mathcal{R}= \lbrace y:|y|<1,|y|<|y-\hat{e}| \rbrace
\eeq
is the rescaled $\mathcal{R}_{12}$.
Integral \reef{eq:A2method} is not convergent when integrating $y$ around $0$ due to the presence of relevant operators being exchanged. These divergences, associated with the renormalization of the operator, need to be subtracted away. This can be again done by computing the integral with a UV cutoff and by dropping terms that diverge when the cutoff goes to zero.

Although Eqs.~\reef{eq:Afinalt} and \reef{eq:A2method} are not manifestly identical, the logic of their derivation shows that they should give identical answers (and they do, in all cases we checked). In practical computations, both ways of proceeding have advantages and disadvantages. Method 2 fully takes advantage of the symmetry among $0,1,\infty$, while the integrands in Method 1 do not respect this symmetry (it is broken by the subtraction terms). Still, if one were to aim for analytic expressions, Method 1 seems preferable. The shape of the integration region in Method 2 makes it hard to compute the integral analytically. However, Method 2 will prove useful and yield more precise results when the integral needs to be evaluated numerically. Besides, in $d=3$, where the correlation function is not known exactly but will be constructed approximately from the bootstrap data, Method 2 allows to consider the conformal block expansion in the s-channel only, without any need to deal with the t- and u-channel decomposition.

We adopted Method 2 as the principal method for the beta-function computation both in $d=2$ and $d=3$, since as we will see the integrals have to be computed numerically. While Method 1 is less precise for the numerical evaluation, we still checked that it gives the same results within its reduced precision. 

\subsection{Beta-function: $d=2$}

The 2d SRFP is the minimal model CFT $\calM(3,4)$ \cite{Belavin:1984vu} and everything about it is known exactly. In particular, we have $\Delta_\sigma=1/8$, $\Delta_\vareps=1$, $C_{\sigma\sigma\vareps}=1/2$. The 4pt function of $\sigma$ is given by 
\beq
\label{eq:4sigma2d}
\langle \sigma(0)\sigma(1)\sigma(z)\sigma(\infty)\rangle =  \frac{|1+\sqrt{1-z}|+|1-\sqrt{1-z}|}{2 |z|^{1/4}|1-z|^{1/4}}\,.
\eeq
Comparing the notation to \reef{eq:Afinalt}, here we have fixed $\hat e$ at $1$ on the real axis, while $y=z$ runs over the full complex plane. In spite of the appearance the 4pt function is smooth across $z\in(1,+\infty)$. 

The 4pt function of $\chi$ is gaussian, given by the sum of three Wick contractions. In the same kinematics,
\beq
\label{eq:4chi2d}
\langle \chi(0)\chi(1)\chi(z)\chi(\infty)\rangle = 1+\frac 1{|z|^{2\Delta_\chi}} +\frac 1{|1-z|^{2\Delta_\chi}}\,,\quad \Delta_\chi=2-\Delta_\sigma=15/8\,.
\eeq
The 4pt function of $\calO$ is given by the product of \reef{eq:4sigma2d} and \reef{eq:4chi2d}:
\begin{align}
\label{eq:F2d}
F(z,\bar z) = \left [1 + \frac{1}{|z|^{\frac{15}{4}}} + \frac{1}{|z - 1|^{\frac{15}{4}}} \right ]\frac{|1 + \sqrt{1 - z}| + |1 - \sqrt{1 - z}|}{2|z|^{\frac{1}{4}}|z - 1|^{\frac{1}{4}}} 
\,.
 \end{align}
We were not able to evaluate the integral of $F(z)$ analytically, so we will report the results of the numerical evaluation. We employ Method 2, so that we have to integrate over the region $\mathcal{R}$.

As discussed after Eq.~\reef{eq:Afinal}, the integral is not convergent around $0$. If we expand $F(z,\bar z)$ around $z=0$, we encounter several terms responsible for the non-convergence. The terms $|z|^{-4}$ and $|z|^{-2}$ correspond to contributions of the identity operator and energy density $\varepsilon$ respectively. Other terms, such as $z/|z|^3$ and $z^2/|z|^4$ ($+$h.c.), are the contributions of $\del_\mu \vareps$ and $T_{\mu\nu}$ in the $\calO\times\calO$ OPE; however, they will vanish upon angular integration.

\begin{figure}
\centering
\includegraphics[scale=1]{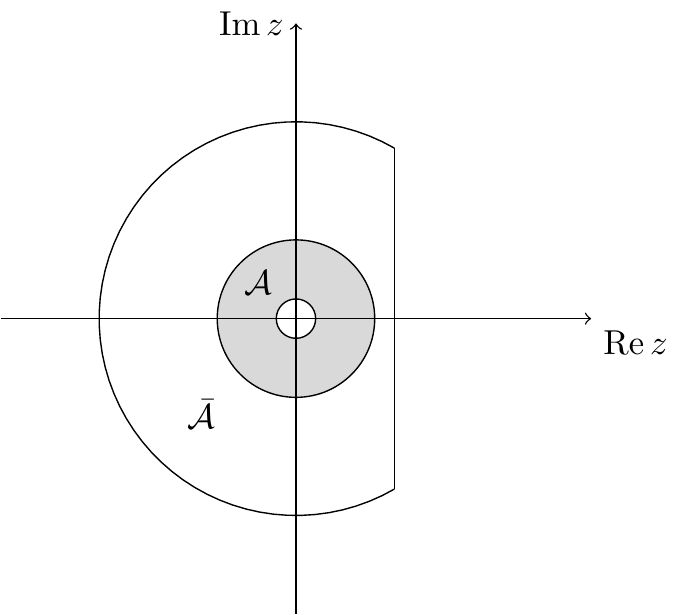}
\caption{The integration region $\mathcal{R}$.}
\label{fig:Rregion}
\end{figure}

To deal with the divergences, we remove from the region $\mathcal{R}$ a small disk $|z|<a$ around the origin, and divide the rest into two regions:
the annulus $\mathcal{A}(a<|z|<r_0$) centered around zero and its complement $\bar{\mathcal{A}}$, see Fig.~\ref{fig:Rregion}. Here $r_0$ is arbitrary subject to $a<r_0<1/2$. In $\mathcal{A}$ we expand $F(z)$ functions as a series in $z$ and $\zb$ up to some high order. We can then drop the terms that vanish upon angular integration, and we integrate exactly the remaining terms. The power-divergent, as $a\to0$, part of the answer is dropped. In the complement of the annulus we integrate $F(z)$ numerically. The so regulated integral over $\calR$ is then:
\beq
I_{d=2}=\int_{\calR} d^2z\, F(z,\bar z) = -0.403746\ldots
\eeq
All the shown digits are exact, and we checked that the result is stable against changes of $r_0$. This implies 
\beq
\label{eq:beta3d=2}
\beta_3= -3(2\pi) I_{d=2}/3!= 1.268404\quad(d=2).
\eeq

\subsection{Beta-function: $d=3$}
The $d=3$ SRFP is not yet exactly solved; however, high precision results are available thanks to the progress of the numerical conformal bootstrap \cite{ElShowk:2012ht,El-Showk:2014dwa,Simmons-Duffin:2015qma,Kos:2016ysd,Simmons-Duffin:2016wlq}. Recently, the approximate critical 3d Ising 4pt function extracted from the bootstrap data was used in \cite{Komargodski:2016auf} to study the random bond Ising critical point. It was also used in \cite{Rychkov:2016mrc} to qualify the non-gaussianity of the 3d Ising model.

Here we proceed analogously and will use the OPE coefficients $C_{\sigma \sigma \calO}$ and dimensions $\Delta_\calO$ of the lowest lying operators (such that $\Delta_\calO$ is smaller than some cutoff in the spectrum $\Delta_*$) to construct an approximate 4pt function for the $\sigma$ field:\beq
\langle \sigma(0) \sigma(\hat e)\sigma(y)\sigma(\infty) \rangle\simeq \frac{1}{|x|^{2 \Delta_\sigma}} \sum_{\calO:\Delta_\calO<\Delta_*}C_{\sigma \sigma \calO}^2\, g_{\Delta_\calO,\l_\calO}(z,\bar z)\,,
\eeq
where $g_{\Delta,\l}$ are the conformal blocks. Let us fix $\hat e=(1,0,\ldots,0)$. Then $z$ is the complex coordinate related to $y$ by
\beq
z= y_1 + i |y_\perp|,\quad y_\perp = (y_2,\ldots,y_d)\,.
\eeq
The 4pt function only depends on $|y_\perp|$ because of rotation invariance around the $x_1$ axis. The usual conformal cross ratios $u$, $v$ are $u=|z|^2$, $v=|1-z|^2$. Instead of $z$, it will be convenient to work with the radial coordinate $\rho$ \cite{Hogervorst:2013sma}
\beq
\rho(z)=\frac{z}{\left(1+\sqrt{1-z} \right)^2}\,.
\eeq
In three dimensions, the conformal blocks are not known exactly; however, they can be computed efficiently as a series in $r$ and $\eta=\cos \theta$, where $\rho=r e^{i\theta}$, using a recursion relation \cite{Kos:2013tga,Costa:2016xah}. The conformal block expansion converges for $r<1$ \cite{Pappadopulo:2012jk}, while in the integration region $\mathcal{R}$ the maximum value of $r$ is $2-\sqrt{3}\simeq 0.27<1$, so our series expansion will converge exponentially fast.

When approximating the 4pt function, we have to take into account three different sources of error:
\begin{enumerate}
\item We do not know the OPE coefficients and the operator dimensions exactly, as they are obtained through the numerical conformal bootstrap. The uncertainty due to this will turn out to be subleading;
\item We compute the conformal blocks as a series expansion in $r$. Here we did it up to order $O(r^{12})$, which provided sufficient accuracy, but it would be straightforward to compute them to a higher order;
\item We know the dimensions and the OPE coefficients of primary operators only up to a dimension $\Delta_*$. The error introduced is of order $r^{\Delta_*}$ \cite{Pappadopulo:2012jk}. We use data from the numerical conformal bootstrap on operators up to dimension $\Delta_*=8$.
\end{enumerate}

We will focus on the last source of error, since it will be the dominant one. The error one introduces when truncating the conformal block expansions of a 4pt function of identical scalars $\sigma$ to some dimension $\Delta_*$ was estimated in \cite{Pappadopulo:2012jk}(see also \cite{Rychkov:2015lca}) to be
\beq
\left | \sum_{\calO: \Delta_\calO \ge \Delta_*} C_{\sigma \sigma \calO}^2\, g_{\Delta_\calO,\l_\calO}(z,\zb) \right | \lesssim \frac{2^{4 \Delta_\sigma}}{\Gamma(4 \Delta_\sigma + 1)} \Delta_*^{4 \Delta_\sigma} |\rho(z)|^{\Delta_*} \label{eq:trunc}\, .
\eeq
This error estimate is essentially optimal for real $0<z<1$, when the 4pt function is in a reflection positive configuration, and all conformal blocks are positive. This corresponds to the configuration with $\eta=1$ in the $\rho$ plane. For configurations with $\eta<1$, conformal blocks decrease in absolute value by unitarity, and hence the same estimate \reef{eq:trunc} applies, although it's no longer optimal. When we integrate the 4pt function over the $\eta$ coordinate, we will not be in a reflection positive configuration, but we will nonetheless bound the truncated operators contribution by its largest possible value, obtained for $\eta=1$. Clearly, the obtained error estimate will be overly conservative, since it does not take into account cancelations due to the varying sign of contributions of operators with spin.

Once we have constructed the approximated 4pt function, we integrate it in the region $\mathcal{R}$.\footnote{In $r$ and $\eta$ coordinates, the region $\mathcal{R}$ is given by $0<r<r_*(|\eta|)$ and $-1<\eta<1$, with $r_*(\eta)=2+\eta-\sqrt{\eta^2+4\eta+3} $.} We follow the procedure outlined in appendix C of \cite{Komargodski:2016auf}: this consists in expanding the 4pt function as a power series in $r$ and $\eta$, then integrating over $r$ and dropping the diverging contributions of the identity and the energy operator. Finally we series-expand again with respect to $\eta$ and we integrate the result exactly.

The data concerning the operator dimensions up to $\Delta_*=8$ and their OPE coefficients can be found in Table 2 of \cite{Simmons-Duffin:2016wlq} (our $C_{\sigma\sigma\calO}=f_{\sigma\sigma\calO}$ given in that table). The OPE coefficients given there are in the normalization for which the small $r$ limit of the conformal block is $g_{\Delta,\l}\simeq\frac{\l!}{(\nu)_\l} (-1)^\l C_\l^{\nu}(\eta) (4 r)^\Delta+\ldots$, where $C_\l^{\nu}$ is a Gegenbauer polynomial, $\nu=\frac{d}{2}-1=1/2$ and $(\nu)_\ell$ is the Pochhammer symbol. 
Using these values, we obtain $I_{d=3}=-1.950 \pm 0.005$. 
The error is dominated by the truncation error, which we estimate by integrating \reef{eq:trunc}.\footnote{For comparison, if we only use operators up to $\Delta_*=6$, we obtain the same central value but with a much larger error estimate: $I_{d=3}=-1.95\pm0.08$. This confirms that the error estimate is overly conservative.} The $g^3$ term of the beta-function is then 
\beq
\label{eq:beta3d=3}
\beta_{3}=12.26\pm 0.03\quad (d=3).
\eeq 

\subsection{Fixed point existence}
\label{sec:existence}
If $0<\delta\ll 1$, the flow that we are studying will reach a fixed point at 
\beq
\label{eq:FP}
g^2 = g_*^2= \delta/\beta_3\,.
\eeq
This fixed point is naturally identified with LRFP. Notice that for our picture to be correct, we must have $\beta_3>0$ (otherwise the fixed point at real $g$ does not exist). The sign of $\beta_3$ was not manifest in the above calculations, since the regulated integrals are not sign-definite.\footnote{As a curiosity we notice that if it were not for the subtraction terms which had to be introduced in the process of disentangling short-distance divergences, then $A$ would be positive, and $\beta_3$ negative.} Still, we have seen that $\beta_3$ is positive in both $d=2$ and $d=3$. This provides a nontrivial check on our picture. 

That $\beta_3>0$ means that the operator $\sigma\chi$ is marginally irrelevant at the crossover. 
The flow at the crossover will be affected by logarithmic corrections to scaling due to $\sigma\chi$.
One must be aware of this fact when interpreting Monte Carlo simulation data in the crossover region.
See the discussion in appendix \ref{sec:priorphys}.

\section{Anomalous dimensions}
\label{sec:anom}

When deforming a CFT with a local perturbation, operators renormalize and acquire anomalous dimensions. Let us recall how these are computed in conformal perturbation theory. As usual, we require observables to be cutoff independent. To find the anomalous dimension of a local operator $\Phi(x)$,  assumed unit-normalized, we look at an observable with one insertion of $\Phi$, $\langle \Phi(0) \Xi \rangle$. Perturbative corrections will be given by
\beq
\frac{g^n}{n!}\int d^d x_1 \ldots d^d x_n \langle \Phi(0) \calO(x_1) \ldots \calO(x_n) \Xi \rangle\,.
\eeq
We regulate the integral by point splitting, with a short distance cutoff $a$, like in section \ref{sec:beta}. There we dealt with the divergences and cutoff dependence which appear when operators $\calO$ approach each other. Those were taken care of by renormalization of the coupling, leading to the nontrivial beta-function. Now we are interested in the additional divergences, in particular the logarithmic ones, which appear when operators $\calO$ collide with $\Phi$.

We define a renormalized operator $\Phi_{\rm R}$, whose correlation functions remain finite in the $a \to 0$ limit. This is related to the bare operator by
\beq
\Phi = Z_\Phi(g,a) \Phi_{\rm R} \,.
\eeq
The anomalous dimension of $\Phi$ will then be given by 
\beq
\gamma_\Phi =-\frac{1}{Z_\Phi} \frac{\partial Z_\Phi}{\partial \log(1/a)}\,.
\eeq

The above discussion was general, but now let us specialize to the flow which interests us, namely SRFP+$\chi$ perturbed by \reef{eq:ourpert}. We are ultimately interested in $\delta>0$ small, but at the leading order we can compute the anomalous dimension for $\delta=0$, when it's related to the log divergence as above. Moreover, order $g$ corrections will vanish thanks to the $\mathbb{Z}_2^\chi$ symmetry of the unperturbed theory, since $\Phi \calO \Phi $ will be odd no matter if $\Phi$ is even or odd, and hence $C_{\Phi  \calO \Phi}=0$. We will therefore be interested in the anomalous dimension to order $g^2$. The computation of this anomalous dimension parallels the beta-function computation. To extract the short-distance divergence giving rise to the cutoff dependence of $\Phi$, we consider the `triple OPE'
\beq
\Phi(0) \calO(x_1) \calO(x_2) \sim h(x_1,x_2)\Phi(0)\,,
\eeq
where $h$ is the 4pt function
\beq
h(x_1,x_2)=\langle \Phi(0) \calO(x_1) \calO(x_2) \Phi(\infty) \rangle\,.
\eeq
If the short-distance logarithmic divergence is 
\beq
\int_V d^d x_1 d^d x_2 \langle \Phi(0) \calO(x_1) \calO(x_2) \Phi(\infty) \rangle = B \log \frac{1}{a} + \ldots
\label{eq:Blog}
\eeq
the renormalized operator will be made cutoff-independent by the choice
\beq
Z_\Phi = 1 + \frac{g^2}{2} B \log \frac{1}{a} + O(g^3)\,.
\eeq
It follows that at the fixed point, where $g=g_*$, the operator $\Phi$ will acquire an anomalous dimension of
\beq 
\gamma_\Phi = - \frac{g_*^2}{2} B + O(g_*^3)\,. \label{eq:andim}
\eeq

As before, we rescale the two integration points $x_1$ and $x_2$ by $|x_1|$. The logarithmic divergence of the integral \reef{eq:Blog} is then
\beq
B={\rm S}_{d}\int d^d y \langle\Phi(0) \calO(y) \calO(\hat{e}) \Phi(\infty)\rangle
\label{eq:Bdiv}\qquad \text{(naive)}.
\eeq
Just like for the beta-function, this ``naive" answer needs to be regulated because of short-distance power divergences which we neglected. 

Excluding the case $\Phi= \calO$, the OPE $\Phi\times \calO$ does not contain the unit operator and the stress tensor. Nor does it contain $\Phi$ since $C_{\Phi\Phi\calO}=0$. Assuming all other operators in the OPE have dimension larger than $\Phi$, the above integral is convergent near $0$ and $\infty$. Let us proceed under the above assumptions, otherwise minor obvious modifications will be required.

The integral does present power divergences for $y$ close to $\hat{e}$. These divergences are due to the unit operator and $\varepsilon$ in the $\calO\times\calO$ OPE. As already mentioned in the beta-function discussion, they do not have anything to do with the critical point physics. We have to subtract and drop these divergences, but we have to do this in a way which does not modify the log divergence influencing the anomalous dimension of the operator $\Phi$. We have again two different ways to proceed, with minor modifications compared to the beta-function computation.

\emph{Method 1.} We subtract the contributions of the relevant operators at the level of Eq.~\reef{eq:Blog}, so that the logarithmic divergences are unchanged. Just as \reef{eq:Bdiv}, \reef{eq:Blog} diverges only when $x_1\to x_2$, while it is finite for $x_i$ close to $0$ and to $\infty$. Given that the relevant operators appearing in the $\calO \times \calO$ OPE are the identity and the $\varepsilon$ operator, we can use the subtraction
\beq
\langle\Phi(0) \calO(x_1) \calO(x_2) \Phi(\infty)\rangle \to \langle\Phi(0) \calO(x_1) \calO(x_2) \Phi(\infty)\rangle - \frac{1}{|x_1-x_2|^{2d}}-\frac{C_{\Phi \Phi \varepsilon}C_{\calO \calO \varepsilon}}{|x_1 - x_2|^{2d-\Delta_\varepsilon}|x_1|^{\Delta_\varepsilon}}\,.
\eeq
Then we rescale the points by $|x_1|$, we obtain the following regulated expression for the logarithmic divergence coefficient:
\beq
B={\rm S}_{d}\int d^d y \left[ \langle\Phi(0) \calO(y) \calO(\hat{e}) \Phi(\infty)\rangle - \frac{1}{|y-\hat{e}|^{2d}}- \frac{C_{\Phi \Phi \varepsilon}C_{\calO \calO \varepsilon}}{|y-\hat{e}|^{2d-\Delta_\varepsilon}} \right]\,.
\label{eq:Bmethod1}
\eeq

\emph{Method 2.} We split again the integration region of \reef{eq:Blog} into three smaller subregion. This will make the numerical evaluation of the integral simpler. Clearly, the contribution of the integration region with $x_1$ close to zero, $|x_1|<|x_2|$ and $|x_1|<|x_1-x_2|$, is the same as that of the region with $x_2$ close to zero. However, the contribution of the region where $x_1$ and $x_2$ are close together will be different. By the same logic as for \reef{eq:A2method}, we obtain a regulated expression for \reef{eq:Bdiv}:
\beq
B={\rm S}_{d} \int_\mathcal{R} d^d y \left\lbrace 2 \langle\Phi(0) \calO(y) \calO(\hat{e}) \Phi(\infty) \rangle + \langle \calO(0) \calO(y) \Phi(\hat{e}) \Phi(\infty) \rangle \right\rbrace\,.
\eeq
The integration region $\mathcal{R}$ is the same as in the previous section. 
The first term is finite since $y$ is separated from $\hat e$. The second term has powerlike divergences, but no log divergences, for $y$ close to $0$; we make it finite by dropping the divergent terms.

\subsection{Results: $d=2$}
\label{sec:results}

We will now apply the developed formalism to determine the anomalous dimension of a few selected operators, first in $d=2$ and then in $d=3$.

\subsubsection{$\chi$, $\sigma$ and $\calO$}
\label{sec:chi-sigma}

The arguments and the results of this section work for any dimension, so we keep $d$ general.
We will see in section \ref{sec:all-order} that the anomalous dimensions of the three operators we consider here can be discussed to all orders. As a check, we will reproduce here the lowest-order versions of those results using the general formalism.

We consider first the field $\chi$. It clearly plays a very special role the $\sigma\chi$-flow, being described by a non-local action in the UV. As a consequence, we expect that $\chi$ does not get anomalous dimension to all orders in $\delta$. This is similar to what happens for the $\phi$ field in the $\phi^4$-flow. Here we will check, by an explicit computation, that the anomalous dimension of $\chi$ vanishes at order $g^2$. 

To see this, observe that the integral \reef{eq:Blog}, with $\Phi=\chi$ and $\calO=\sigma\chi$, only has power divergences, and no logarithmic divergences. Indeed its integrand is 
\beq
\frac{1}{|x_1-x_2|^{2\Delta_\sigma}}\left(
\frac{1}{|x_1|^{2\Delta_\chi}}+
\frac{1}{|x_1|^{2\Delta_\chi}}+
\frac{1}{|x_1-x_2|^{2\Delta_\chi}}\right)\,.
\eeq
The integral only has power divergences by the same argument as that given for the beta-function subtraction terms \reef{eq:subterms}. If we were to apply Method 1 to this integral, we would end up with an identically vanishing integrand. Notice that in this case the OPE $\Phi\times \calO$ contains an operator $\sigma$ with dimension $\Delta_\sigma<\Delta_\Phi$, So more subtraction terms are needed than the ones given in Eq.~\reef{eq:Bmethod1}. After these subtractions, the integrand is identically zero.

Next we consider the field $\sigma$. It is also special, because it acts as a source for $\chi$, and so the classical equation of motion (EOM) of $\chi$ sets a linear non-local relation between the two. In quantum theory, this non-local EOM implies that the IR dimensions of $\sigma$ and $\chi$ satisfy the shadow relation:
\beq
\Delta_\chi+\Delta_\sigma=d\,.
\eeq
This should be compared with the shadow relation \reef{sec:shadow0} for the $\phi^4$-flow. The two relations suggest that in the IR limit of the two flows, $\phi$ has to be identified with $\sigma$, and $\phi^3$ with $\chi$. This fits nicely our proposed IR duality and will be discussed further in section \ref{sec:dual}.

Here we will check the shadow relation at the leading order in $g$. The anomalous dimension of the spin field $\sigma$ can be reduced by a trick to the $g^3$ term of the beta-function, which we already computed. Let us consider the original integral \reef{eq:Blog} for $\Phi=\sigma$. It's easy to see that this integral (multiplied by the overall volume) is exactly one third of the integral \reef{eq:intf} in the beta-function calculation. Indeed, the integrand in both cases involves the 4pt function of $\sigma$ multiplied by a correlation function of $\chi$, which has one term in the first case and three terms in the second one. These three terms all contribute equally, and so we obtain $B=A/3$. This fact is not manifest the expressions provided by Methods 1 and 2, but we checked it numerically. Given $B=A/3$, the anomalous dimension of $\sigma$ is found to be:
\beq
\gamma_\sigma= \delta+ O(\delta^2)\,.
\eeq
This checks the shadow relation at the lowest order. 

{We pause to notice that, as a consequence of the discussed RG equations, the 2pt function $\langle\sigma\sigma\rangle$ \emph{at the crossover} ($\delta=0$) exhibits a $1/\log r$ suppression in presence of the marginally irrelevant $\sigma\chi$ perturbation. See appendix \ref{sec:log}.}

Finally, we discuss the operator $\calO=\sigma\chi$ which drives the flow. By the general RG arguments, the anomalous dimension of this operator should be given by:
\beq
\gamma_\calO(g)=\beta'(g)\,.
\eeq
Using the leading beta-function expression, at the fixed point this becomes
\beq
\label{eq:gammaO}
\gamma_\calO(g_*)=3\delta\,,
\eeq
up to order $g^2$ corrections. As expected, $\calO$ becomes irrelevant at the IR fixed point:
\beq
[\calO]_{\rm LRFP}=d+2\delta+O(\delta^2),
\eeq
Eq.~\reef{eq:gammaO} can also be obtained in the formalism of the previous section. It follows by noticing that $B=A$ for the renormalization of $\calO$.

\subsubsection{$\varepsilon$}

To compute the anomalous dimension of $\varepsilon$, we need to use the result from the Ising minimal model (see e.g.~\cite{Mattis:1986mj})
\beq
\label{eq:Mattis}
\langle \varepsilon(0) \sigma(z) \sigma(1) \varepsilon(\infty) \rangle =\frac{|1+z|^2}{4 |z||1-z|^{1/4}}\,.
\eeq
To obtain the correlation function $\langle \varepsilon \calO \calO \varepsilon\rangle$ we multiply \reef{eq:Mattis} by $\langle \chi(z)\chi(1)\rangle$. 
This time we will use Method 1, and we will be able to carry out the integration analytically. We need to subtract the divergent terms due to the relevant operators, as shown in \reef{eq:Bmethod1}. The $\varepsilon$ subtraction term is absent since, thanks to the Kramers-Wannier duality, $C_{\varepsilon \varepsilon \varepsilon}=0$ in two dimensions.
We obtain the integral
\beq
\int d^2 z \frac{1}{|1-z|^{4}}\left(\frac{|1+z|^2}{4 |z|}-1\right)\,, \label{eq:epsInt}
\eeq
to be computed with circular cutoffs around $0$, $1$ and infinity. Careful evaluation shows that this integral is zero, see appendix \ref{sec:integral} for the proof. 
Unfortunately, the only proof we found was by brute force, and it would be nice to find an underlying reason. We also checked this result by numerical evaluation. Numerically, this was also previously observed in \cite{Komargodski:2016auf}, Eqs.~(6.14) and (5.28), in an unrelated computation which led to the same integral.

Therefore the anomalous dimension of $\varepsilon$ vanishes at order $g_*^2$, while order $g_*^3$ will be zero by the $\bZ_2$ selection rules. Therefore
\beq
\label{eq:gammaepsd=2}
\gamma_\varepsilon = O(g_*^4)=O(\delta^2)\qquad (d=2)\,.
\eeq

\subsubsection{$T_{\mu\nu}$}
We now come to the discussion of the stress tensor operator, the source of some paradoxes discussed in the introduction. The LRFP is a non-local theory, and we do not expect it to contain a conserved local stress tensor operator. Let us examine this issue from the RG point of view. The UV theory SRFP+$\chi$ consists of two decoupled sectors. The SRFP is a local theory, with a conserved local stress tensor which we call $T_{\mu\nu}$. The $\chi$ sector is non-local, without a local stress tensor.\footnote{This is easy to check explicitly. The $\chi$ sector being gaussian, all local operators are normal-ordered products of $\chi$ and its derivatives, and by inspection there is no spin 2, dimension $d$ operator which could play the role of a local stress tensor.} When we perturb the UV theory with the operator $\sigma \chi$, the two sectors are no longer decoupled and locality of the SRFP is lost. This implies that, at the IR fixed point, the operator $T_{\mu\nu}$ will acquire an anomalous dimension. We will still call it $T_{\mu\nu}$ and will sometimes refer to it as the `stress tensor', but it has to be kept in mind that this operator will not be conserved at the IR fixed point.

We will compute the $T_{\mu\nu}$ anomalous dimension in two ways, first directly and then using the multiplet recombination which will clarify the puzzle of missing states.

For the direct computation, it is sufficient to consider only one tensor component, say $T \equiv T_{zz}$, as all the components will acquire the same anomalous dimension. The stress tensor in the UV is conventionally normalized as
\beq
\langle T(z,\bar{z}) T(0)\rangle = \frac{c}{2} \frac{1}{z^4}\,.
\eeq 
In the case of the two dimensional Ising model, the central charge is $c=\frac{1}{2}$. The 4pt function $\langle T(0) \sigma(z) \sigma(1) T(\infty) \rangle$ is then recovered in the standard way using the Ward identity twice on the 2pt function of $\sigma$. For the 4pt function involving two $\calO$ insertions we obtain:
\beq
\langle T(0)\sigma\chi(z)\sigma\chi(1)T(\infty) \rangle=\frac{1}{|1-z|^{4}}\left(\frac{1}{4}+\frac{(1-z)^{2}(z^{2}+30z+1)}{256z^{2}}\right)\,.
\eeq
Although the stress tensor is not a scalar operator, the discussion of section \ref{sec:anom} on how to compute the anomalous dimensions still applies. We aim for an analytic result and use Method 1. Since $C_{T T \varepsilon}=0$ in $d=2$, we only need to subtract the contribution of the identity in \reef{eq:Bmethod1}. Note that since the stress tensor is not unit-normalized, subtracting the contribution of the identity means subtracting $(4|z-1|^4)^{-1}$. The resulting integral can be evaluated exactly:
\beq
\int d^2 z \frac{1}{(1-\bar{z})^{2}}\frac{(z^{2}+30z+1)}{256z^{2}}=-\frac{15}{128}\pi\,. \label{eq:TInt}
\eeq
There is a subtlety in this computation related to the contribution of the region near $z=1$. This is explained in appendix \ref{sec:integral}.

Eq.~\reef{eq:andim} as written is valid for the unit-normalized operators. To make up for the fact that $T$ is not, we need to multiply its r.h.s.~by an extra factor $2/c$. Finally, we obtain:
\beq
\gamma_T=\frac{15}{32}\pi^2 g_*^2+O(g_*^4)\approx 3.65\, \delta+O(\delta^2)\quad(d=2), \label{eq:gammaT1}
\eeq
where we used \reef{eq:FP} and \reef{eq:beta3d=2}.

We will now recompute the same anomalous dimension using the recombination of multiplets.\footnote{For recent discussions of multiplet recombination in various CFT contexts see e.g.~\cite{Rychkov:2015naa,Skvortsov:2015pea,Giombi:2016hkj,Roumpedakis:2016qcg}.} As we will see, this method requires only the integration of a 3pt function fixed by a Ward identity, and gives $\gamma_T$ as a function of $\Delta_\sigma$ and of the central charge $c$ for arbitrary $d$. So we switch to general $d$ until the end of this section.

The stress tensor at the SRFP satisfies conservation equation $\del^\mu T_{\mu \nu}=0$, meaning that some of his descendants are zero. As we say, it belongs to a short multiplet. The same operator taken to the IR, to the LRFP, which is a non-local theory, is not expected to be conserved: $\del^\mu T_{\mu \nu} \propto V_\nu\ne 0$. In other words, the stress tensor multiplet becomes long by eating the $V_\nu$ multiplet. The vector $V_\nu$ must exist in the UV theory as well; this was puzzling in the standard picture. The puzzle is neatly resolved in our picture, since this multiplet can be easily constructed with the help of the $\chi$ field. Namely, we have:
\beq
V_{\nu} = \sigma (\del_\nu \chi) -\frac{\Delta_\chi}{\Delta_\sigma}(\del_\nu \sigma)\chi\,.
\eeq
This is clearly a vector field and of dimension $d+1$ at the crossover point. The relative coefficient between the two terms is fixed by requiring that $V_\nu$ be a (non-unit normalized) vector primary at the crossover. For this it is sufficient to check that the 2pt function of $V_\nu$ and of the descendant $\del_\nu(\sigma\chi)$ vanishes.

Since $V_\mu$ given above is the only candidate to be eaten, at the IR fixed point we expect 
\beq
\label{eq:recomb}
\del^\mu T_{\mu \nu}=b(g) V_\nu\,,
\eeq 
where $b(g) \to 0$ as $g \to 0$. 

We will be interested in the first nontrivial order: $b(g) = b_1 g+O(g^2)$. The value of $b_1$ can be determined by studying the 2pt function of $V_{\mu}$ with $T_{\mu \nu}$, computed at first order in perturbation theory. It will be more convenient to utilise the descendant $\del^\mu T_{\mu \nu}$, as this will allow us to use the Ward identity. On the one hand from multiplet recombination \reef{eq:recomb} we expect at the lowest order in $g_*$:
\beq
\langle \del^\mu T_{\mu \nu}(x) V_{\rho}(y) \rangle_g \approx b_1 g_* \langle V_{\nu}(x) V_{\rho}(y)\rangle_0\,. \label{eq:geta}
\eeq
Here and below we mark with subscript $g$ the IR fixed point correlators, while with subscript 0 the correlators in the UV theory SRFP+$\chi$. 
The 2pt function of $V_\mu$ entering this equation can be computed explicitly given its definition:
\beq
\langle V_{\mu}(x) V_{\nu}(0) \rangle_0=2d \frac{\Delta_\chi}{\Delta_\sigma}\frac{I_{\mu\nu}(x)}{|x|^{2d+2}},\quad
I_{\mu \nu}(x)=\delta_{\mu \nu} -2\frac{x_\mu x_\nu}{x^2}\,.  \label{eq:normV}
\eeq
Notice that this functional form is consistent with the conformal primary nature of $V_\mu$.

On the other hand perturbation theory predicts for the correlator in the l.h.s.~of \reef{eq:geta}
\beq
\langle \del^\mu T_{\mu \nu}(x) V_{\rho}(y) \rangle_g = g_* \int d^d z \langle \del^\mu T_{\mu \nu}(x) V_{\rho}(y) \calO(z)\rangle_0\,.  \label{eq:geta1}
\eeq
The 3pt function that we need to integrate is the sum of two factorized terms:
\begin{multline}
\langle \del^\mu T_{\mu \nu}(x) V_{\rho}(y) \calO(z) \rangle_0=\langle \del^\mu T_{\mu \nu}(x) \sigma (y) \sigma(z)\rangle \langle \del_\rho \chi(y) \chi(z) \rangle_0 \\ - \frac{\Delta_\chi}{\Delta_\sigma}\langle \del^\mu T_{\mu \nu}(x)\del_\rho \sigma (y) \sigma(z)\rangle \langle \chi(y) \chi(z) \rangle_0\,.
\end{multline}
When the 3pt functions in the r.h.s.~are expressed using the Ward identity\footnote{\label{note:Ward}In this general $d$ argument we normalize the stress tensor so that the Ward identity takes the form
$\label{eq:Ward}
\langle \del^\mu T_{\mu \nu}(x) \calO_1(x_1)\ldots \calO_n(x_n) \rangle=-\sum_i \delta \left(x-x_i\right) \del_\nu^{x_i} \langle \calO_1(x_1) \ldots \calO_n(x_n) \rangle$\,. Notice that it's not the same as the normalization usually used in 2d.}
of the unperturbed theory, we get terms proportional to $\delta(x-y)$ and to $\delta(x-z)$. We assume that $x \ne y$, so only $\propto\delta(x-z)$ terms are important. They yield a non-zero contribution when we integrate over $z$. We obtain
\begin{multline}
\int d^d z \langle \del^\mu T_{\mu \nu}(x) V_{\rho}(0) \calO(z)\rangle_0
\\
=-\langle \sigma(x) \del_\nu \sigma(0) \rangle \langle \del_\rho \chi(x)\chi(0) \rangle+\frac{\Delta_\chi}{\Delta_\sigma} \langle \del_\nu \sigma(x) \del_\rho \sigma(0) \rangle \langle \chi(x) \chi(0)\rangle 
=2\Delta_\chi \frac{I_{\nu \rho}(x)}{|x|^{2d+2}}\,.
\end{multline}
Using \reef{eq:geta}, \reef{eq:normV}, \reef{eq:geta1}, the value of $b_1$ is fixed:
\beq
b_1=\Delta_\sigma/d\,.
\eeq

Now let us compute the anomalous dimension of $T_{\mu\nu}$. 
The 2pt function normalization customary for $d$ dimensional CFT is \cite{Osborn:1993cr} (see also \cite{Rattazzi:2010gj})
\beq
\label{eq:TTSRFP}
\langle T_{\mu \nu}(x) T_{\rho \sigma}(0) \rangle = \frac{c_T}{2{\rm S}_{d}^2} \frac{1}{|x|^{2\Delta_T}} \left[I_{\mu \rho}(x) I_{\nu \sigma}(x)+\left(\mu \leftrightarrow \nu \right) -\frac{2}{d}\delta_{\mu \nu} \delta_{\lambda \sigma}\right]\,,
\eeq
In this normalization, and assuming the Ward identities are normalized as in note \ref{note:Ward}, the free massless scalar has $c_T=d/(d-1)$.

Eq.~\reef{eq:TTSRFP} follows just from conformal invariance and the fact that $T_{\mu\nu}$ transforms as a rank 2 symmetric traceless primary.
So it's valid both at the SRFP in the UV, and at the LRFP in the IR. In this argument we assumed conformal invariance of the LRFP, which will be discussed below in section \ref{sec:standard}.\footnote{It's also possible to see without invoking conformal invariance that the tensor structure of the 2pt function is preserved along the RG flow. This follows from the fact that the rescaling needed to make the operator finite depends only on the indices of the operator and not on any other insertions in the correlation function. It's part of the same argument which shows that all tensor components get the same anomalous dimension.} In the UV we have $c_T=c_T^{\rm SRFP}$ and $\Delta_T=d$, corresponding to the conserved local stress tensor. In the IR both $c_T$ and $\Delta_T$ receive $O(g_*^2)$ corrections.  At the intermediate distances there is some interpolating behavior which will not be important. 

%
Let $\Delta_T=d+\gamma_T$ in the IR, where $\gamma_T$ is the anomalous dimension. The quantity of interest is the 2pt function of the divergence of $T_{\mu\nu}$ at the LRFP which can be found by an explicit differentiation of \reef{eq:TTSRFP}. This vanishes for $\gamma_T=0$, consistent with the fact that $T_{\mu\nu}$ is conserved in the UV, and for nonzero $\gamma_T$ is given by:
\beq
\langle \del^\mu T_{\mu \nu}(x) \del^\rho T_{\rho \sigma}(0) \rangle_{g} \approx \frac{c_T}{{\rm S}_{d}^2} \gamma_T \left(d+1-\frac{2}{d}\right) \frac{I_{\nu \sigma }}{|x|^{2d+2}}\,, \label{eq:rec1}
\eeq
In \reef{eq:rec1} we dropped terms higher order in $g_*^2$. One such higher order term is the correction to $c_T$ which will not play any role, so in all subsequent equations $c_T=c_T^{\rm SRFP}$. 

At the same order, using the recombination of multiplets equation~\reef{eq:recomb}, we expect:
\beq
\langle \del^\mu T_{\mu \nu}(x) \del^\rho T_{\rho \sigma}(0) \rangle_{g}\approx b_1^2 g_*^2\langle V_{\nu}(x) V_\sigma(0)\rangle_{0}\,.\label{eq:rec2}
\eeq
From the last two equations, the 2pt function of $V_\mu$, and the value of $b_1$ we find the lowest-order anomalous dimension of the stress tensor:
\beq
\boxed{\gamma_T=\frac{2{\rm S}_{d}^2}{c_T}\frac{\Delta_\sigma (d-\Delta_\sigma)}{d^2+d-2} g_*^2+O(g_*^4)}\,. \label{eq:gammaT2}
\eeq

Let us now specialize to $d=2$. In the usual 2d normalization, the 2d critical Ising has central charge $1/2$, half that of the free massless scalar. As mentioned, $c_T=\frac{d}{d-1}=2$ for the free massless scalar in the normalization of \reef{eq:TTSRFP} and of note \ref{note:Ward}, and so $c_T=1$ for the 2d Ising in the same normalization. It is then easy to see that \reef{eq:gammaT2} agrees with the result \reef{eq:gammaT1} obtained via the integration of the 4pt function.

\subsection{Results: $d=3$}

The anomalous dimensions of $\chi$, $\sigma$ and $\calO$ were already discussed in section \ref{sec:chi-sigma} for any $d$.

\subsubsection{$\vareps$}

Recall that order $g_*^2$ anomalous dimension of the energy operator was zero in $d=2$, for mysterious reasons unexplained by any obvious symmetry. As we will see this does not happen in three dimensions. We set up a numerical computation for this anomalous dimension using the CFT data from the numerical conformal bootstrap. In order to compute the 4pt function $\langle \varepsilon \sigma \sigma \varepsilon \rangle$, we will need the operator dimensions and the OPE coefficients of the operators appearing in the $\sigma \times \sigma$, $\sigma \times \varepsilon$ and $\varepsilon \times \varepsilon$ OPEs. For operators up to $\Delta_*=8$, these can be found in Table 2 of \cite{Simmons-Duffin:2016wlq}. We will use Method 2. We construct the 4pt function in the region where one $\varepsilon$ is close to one $\calO$ and the region where the $\calO$'s are close together:
\begin{gather}
\langle\varepsilon(0)\sigma(z)\sigma(1)\varepsilon(\infty)\rangle=\frac{1}{|z|^{\Delta_\sigma+\Delta_\varepsilon}}\sum_{\calO:\Delta_\calO<\Delta_*}C_{\sigma \varepsilon \calO}^2 g^{\Delta_{\vareps \sigma},\Delta_{\sigma \vareps}}_{\Delta_\calO,\l_\calO}(z,\bar{z})\,, \label{eq:eps3d1}\\
\langle\sigma(0)\sigma(z)\varepsilon(1)\varepsilon(\infty)\rangle=\frac{1}{|z|^{2\Delta_\sigma}}\sum_{\calO:\Delta_\calO<\Delta_*}C_{\sigma \sigma \calO} C_{\varepsilon \varepsilon \calO} g^{0,0}_{\Delta_\calO,\l_\calO}(z,\bar{z})\,. \label{eq:eps3d2}
\end{gather}
Here $g_{\Delta_\calO,\l_\calO}$ with upper indices are the conformal blocks for the external scalars with unequal dimensions, which we compute via recursion relations from \cite{Kos:2014bka}. The operators entering the sum in the first (resp.~second) equation are $\bZ_2$ odd (resp.~even). In both cases, we will be integrating over the region $\mathcal{R}$ defined in section \ref{sec:beta}.

Once again, the largest error contribution when approximating the 4pt function will come from the truncation of the spectrum at dimension $\Delta_*$. The same line of reasoning used to obtain the truncation error for four identical scalar in \cite{Pappadopulo:2012jk} will go through in the case of equation \reef{eq:eps3d1}. Analyzing the proof in \cite{Pappadopulo:2012jk}, it's possible to see that the truncation error will be given by \reef{eq:trunc} with the change $\Delta_\sigma \to (\Delta_\sigma+\Delta_\vareps)/2$ in all occurrences in the r.h.s.

For equation \reef{eq:eps3d2}, however, we cannot map the 4pt function onto a reflection positive configuration, and therefore we cannot find a bound on the contribution of the truncated operators in the same way. We need to first use Cauchy's inequality so that the tail of $\langle \sigma \sigma \varepsilon \varepsilon \rangle$ can be bounded by the tails of $\langle \sigma \sigma \sigma \sigma \rangle$ and $\langle \varepsilon \varepsilon \varepsilon \varepsilon \rangle$. At this point we can use again the result of \cite{Pappadopulo:2012jk}, and we obtain
\beq
\left| \sum_{\calO:\Delta_\calO>\Delta_*}C_{\sigma \sigma \calO} C_{\varepsilon \varepsilon \calO} g^{0,0}_{\Delta_\calO,\l_\calO}(z,\bar{z})\right| \lesssim \frac{2^{2\Delta_\sigma +2 \Delta_\varepsilon}}{\sqrt{\Gamma(4\Delta_\sigma+1) \Gamma(4\Delta_\varepsilon+1)}} \Delta_*^{2\Delta_\sigma +2 \Delta_\varepsilon} |\rho(z)|^{\Delta_*}\,.
\eeq

Truncating the CFT data up to $\Delta_*=8$,\footnote{Beware of the changes in normalization of OPE coefficients between \cite{Simmons-Duffin:2016wlq} and \cite{Kos:2014bka,Komargodski:2016auf}, explained in appendix A.3 in \cite{Simmons-Duffin:2016wlq}.} and carrying out the integration in the region $\mathcal{R}$, we obtain a nonzero value, unlike in $d=2$. The order $g_*^2$ anomalous dimension is
\beq
\label{eq:gammaeps}
\gamma_\varepsilon\approx 3.3 g_*^2+O(g_*^4)\approx 0.27\,\delta+O(\delta^2)\quad (d=3),
\eeq
where in the second equality we used \reef{eq:FP} and \reef{eq:beta3d=3}.
The total truncation error on the coefficient 3.3, estimated as above, is $\pm 0.5$.
So we are confident that $\varepsilon$ gets a nonzero anomalous dimension in $d=3$ already at the lowest order allowed by the $\bZ_2$ selection rules.

\subsubsection{$T_{\mu\nu}$}

In $d=3$ the data needed to compute the 4pt function $\langle T \sigma\sigma T \rangle$ are not yet available. So we cannot compute the anomalous dimension of $T_{\mu\nu}$ using the formalism of section \ref{sec:anom}. However, we can still use the general $d$ expression \reef{eq:gammaT2} obtained by using the recombination method.

Using the spin field dimension $\Delta_\sigma = 0.5181489(10)$ \cite{Kos:2016ysd} and the central charge $c_T/c_T^{\rm free}=0.946539(1)$
\cite{Komargodski:2016auf,El-Showk:2014dwa} with the free scalar central charge $c_T^{\rm free}=d/(d-1)=3/2$, we get
\beq
\gamma_T=28.60555(6)g_*^2+O(g_*^4)\approx 2.33\, \delta+O(\delta^2)\,\quad(d=3)\,.
\eeq

\section{Review of the $\phi^4$-flow results}
\label{sec:standard}

Up to now we were focussing on the novel $\sigma\chi$-flow, in the vicinity of $s=s_*$ where it is weakly coupled. As mentioned in the introduction, we propose an infrared duality of the $\sigma\chi$-flow and the $\phi^4$-flow. As a preparation for the discussion of duality, here we will review what is known about the $\phi^4$-flow.

The $\phi^4$-flow is a time-honored tool in the study of the LRFP, whose use goes back to Fisher et al \cite{Fisher:1972zz}. It is weakly coupled close to $s=d/2$, providing quantitative information about the LRFP in that region. In particular, anomalous dimensions can be computed in a power series expansion in $\eps=2s-d$. For example, operator $\phi^n$ gets anomalous dimension
\beq
\gamma(\phi^n)=[n(n-1)/6]\eps+O(\eps^2)\,.
\eeq
To the shown order this is the same answer as for the Wilson-Fisher fixed point in $4-\eps$ dimensions, but this is largely a one-loop coincidence. For example, the anomalous dimension of $\phi$ remains identically zero at the LRFP while it gets an $O(\eps^2)$ contribution at the Wilson-Fisher fixed point. As we will see, this and some other structural properties of the LRFP can be proven to all orders in perturbation theory in $\eps$. 

The results of this section are valid also for $d=1$, since they are independent of the nature of the long-range to short-range crossover (see note \ref{note:d=1}). Indeed, the long-range Ising model in $d=1$ has a second-order phase transition for $0<s<1$, which is mean-field for $s<1/2$ and non-gaussian for $1/2<s<1$. 

\subsection{Absence of anomalous dimension for $\phi$}

We will be following mostly recent work \cite{Paulos:2015jfa} by some of us, where further references and details can be found. For this discussion we normalize the action and the coupling as
\beq
S=\frac{\calN_s}2 \int d^dx\, \phi\ (-\del^2)^{s/2} \phi+\frac{g_0}{4!}  \int d^d x\, \phi^4\,,
\label{eq:S}
\eeq
choosing $\calN_s$ so that the free 2pt function of $\phi$ is unit-normalized.

All-order statements are proven by setting up perturbation theory within the analytic regularization scheme, where one defines the correlation functions via analytic continuation of position-space Feynman integrals in $\eps$. This automatically subtracts UV divergences. Also this scheme is an example of a mass independent scheme, implying that the mass term does not renormalize and the IR fixed point is reached for zero UV mass. For this reason we dropped the $\phi^2$ term from the original action \reef{standardflow}. 

Although there are no UV divergences, some diagrams contain poles in $\eps$. It helps to reorganize perturbation theory by absorbing these poles into a coupling redefinition. Namely, one defines a dimensionless ``renormalized coupling" $g$ related to the original coupling $g_0$ by 
\beq
\label{eq:g0}
g_0= Z_g(g,\eps)g\mu^\eps
\eeq
where $\mu$ is an energy scale, and $Z_g$ contains a series of ascending poles in $\eps$:
\beq
Z_g(g,\eps)=1+\sum _{k=1}^\infty \frac{f_k(g)}{\eps^k}\,.
\eeq
The factor $Z_g$ is chosen in such a way that perturbative expansion of correlation functions at fixed distances and for a fixed $g$ and $\mu$ no longer contains any poles in $\eps$. That this can be done means that the theory at $\eps=0$ is renormalizable. Renormalizability can be proven similarly to how it's done for the local theories with marginal couplings (see note 6 in \cite{Paulos:2015jfa} for references). Our case is even simpler than the local case, in that there is no wavefunction renormalization, due to the fact that the kinetic term for $\phi$ is non-local. This fact implies that the gamma-function $\gamma_\phi(g)$ vanishes. We now review the standard argument that this also implies vanishing of the anomalous dimension at the fixed point.\footnote{See also note \ref{note:defect} below for a nonperturbative discussion. It can also be shown rigorously from the lattice formulation of the long-range Ising model that the anomalous dimension of $\phi$, if any, is nonnegative. This follows from reflection positivity using the so-called infrared bound, see \cite{aizenmanCritical1988}. We thank Michael Aizenman for pointing this out to us.}

Since $\gamma_\phi(g)=0$, the Callan-Symanzik (CS) equation for the 2pt function of $\phi$ takes a simplified form: 
\beq
\label{CS}
[r\,\del_r+\beta(g) \del_g ]F(r,g)=0,\qquad \langle\phi(x)\phi(0)\rangle=F(r,g)/|x|^{2\Delta_\phi},\qquad r=\mu|x|\,.
\eeq
The beta-function is 
\beq
\beta(g)=\mu\, \del_\mu g|_{g_0}=-\frac{\eps g}{1+g(\log Z_g)'}=-\eps g+g^2 f_1'(g)
\eeq
(for deriving the last equation the absence of poles in $\eps$ is crucial). Integrating the CS equation we find
\beq
\label{Fsq}
F(r,g)=\hat F (\bar g(r,g))
\eeq
where $\bar g(r,g)$ is the ``running coupling" obtained by 
solving the beta-function equation with a boundary condition at $r=1$:
\beq
\label{eq:running}
r\,\del_r g=-\beta(\bar g),\quad \bar g(1,g)=g\,.
\eeq

Note that we are not interested in the limit $\eps\to0$, but in the case when $\eps$ is finite and small. So the reorganization of perturbation theory in terms of the coupling $g$ is not a physical necessity, but a technical tool which disentangles the structure of the perturbative expansion. In principle, the correlation functions obtained by solving the CS equations could be obtain by taking the original perturbation theory in $g_0$ and carefully resumming terms enhanced by poles in $\eps$.  The RG technology is nothing but a systematic way to perform such resummations. See \cite{Vasilev:2004yr}, p.90, for a nice discussion of this point.

The beta-function at the lowest non-trivial order takes the form
\beq
\label{eq:beta-stan}
\beta(g)=-\eps g+K g^2+\ldots,\quad K=(3/2)\Sd\,,
\eeq
so that the IR fixed point is at $g_*= \eps/K$. From \reef{Fsq} we conclude that in the IR, where $\bar g\to g_*$, the 2pt function of $\phi$ behaves as
\beq
\hat F(g_*)/|x|^{2\Delta_\phi}\,.
\eeq
with the same power law as in the UV (although a different constant prefactor). 

The function $\hat F$ can be found by matching with fixed order perturbation theory.  At the lowest nontrivial order we have
(see \cite{Paulos:2015jfa}, Eq.~(3.26))
\beq
\textstyle
\hat F(g)=1+ Q g^2+\ldots,\qquad Q=(\pi^d/6) {\Gamma\left(-\frac d4\right)}/{\Gamma\left(\frac{3d}{4}\right)}<0\,.
\label{eq:Q}
\eeq
This will be useful in section \ref{sec:relope}.

\subsection{Shadow relation between $\phi$ and $\phi^3$}

We have a non-local equation of motion (EOM) which relates the $\phi$ and $\phi^3$ operators:
\beq
\label{eq:NLEOM}
\calN_s (-\del^2)^{s/2}\phi+\frac{g_0^2}{3!} \phi^3 = 0\,.
\eeq
 This equation implies that in the IR the dimensions of $\phi$ and $\phi^3$ should be related by
 \beq
 [\phi^3]=[\phi]+s
 \eeq
 which, given that the IR dimension of $[\phi]$ is known, can be rewritten as the ``shadow relation"
 \beq
 \label{eq:sh}
 [\phi]+[\phi^3]=d\,.
 \eeq

A formal proof of the shadow relation can be given as follows (see \cite{Paulos:2015jfa} for another proof). In perturbation theory, the non-local EOM \reef{eq:NLEOM} simply means that the diagrams contributing to correlation functions of $\phi^3$ are the diagrams of $\phi$ amputated by one propagator.\footnote{\label{note:subtle}This amputation relation is true for the diagrams which connect $\phi$ to an interaction vertex. Diagrams which connect $\phi$ directly to another $\phi$ in the correlator do not have a counterpart for the correlators of $\phi^3$. The non-local EOM maps such diagrams into local terms, and is thus valid modulo such local terms. These local terms are not important for the discussion in this section, but it will be important to keep track of them in section \ref{sec:relope}.}
In other words correlators of $\phi^3$ are related to those of $\phi$ by (in momentum space)
\beq
\frac{g_0}{3!} \langle\phi^3(p)\ldots\rangle = |p|^s \langle\phi(p)\ldots\rangle\,.
\eeq
This can be rewritten as 
\beq
\label{eq:phi3ren}
\frac{g \mu^\eps}{3!} \langle (\phi^3)_{\rm R}(p)\ldots\rangle = |p|^s \langle\phi(p)\ldots\rangle\,.
\eeq
where we expressed $g_0$ via $g$ via \reef{eq:g0} and defined the renormalized $\phi^3$ operator via
\beq
\phi^3 = Z_{\phi^3}\cdot (\phi^3)_{\rm R}\,,\qquad Z_{\phi^3} = Z_g^{-1}\,.
\eeq
Eq.~\reef{eq:phi3ren} implies that the correlation functions of the so defined $(\phi^3)_{\rm R}$ are free of poles in $\eps$. Since $Z_{\phi^3}$ is related to $Z_g^{-1}$, a short computation allows to express $\gamma_{\phi^3}$ via the beta-function all along the flow:
\beq
\gamma_{\phi^3}(g)=\mu\,\del_\mu \log Z_{\phi^3}|_{g_0} = \eps+\beta(g)/g\,.
\eeq
In particular in the IR we have $\gamma_{\phi^3}\to\eps$, which given the dimension of $\phi$ is equivalent to the shadow relation \reef{eq:sh}. This argument proves the shadow relation to all orders in perturbation theory.

\subsection{Conformal invariance of the LRFP}

\label{sec:confinv}

It has been shown in \cite{Paulos:2015jfa} that the IR fixed point of the $\phi^4$-flow has $SO(d+1,1)$ conformal invariance to all orders in perturbation theory. Here we will present a somewhat streamlined version of the same argument. 

Let $G(x_1\ldots x_N)$ be an $N$-point correlation function of $\phi$. We would like to show that at large distances it is conformally invariant. Large here means compared to the distance scale $\mu_c^{-1}$ where, roughly, the RG flow transitions from the UV to IR behavior. For example, we can define the energy scale $\mu_c$ as where the running coupling is half way between the UV and IR fixed points.

The first step is to show that correlation functions of the $\phi^4$-flow satisfy broken conformal Ward identities (for dilatations and special conformal transformations). These identities are valid at all distances and take the form:
\begin{gather}
\sum _{i=1}^N [x_i.\del_{x_i} +\Delta_\phi] G(x_1\ldots x_N)  = B_D=b_D \beta(g)\,\\
\sum _{i=1}^N \left[(2x_i^\mu x_i^\lambda-\delta^{\mu\lambda}x_i^2)\frac{\del}{\del{x^\mu_i}} +2\Delta_\phi x_i^\lambda\right]G(x_1\ldots x_N) = B_k = b_K \beta(g)\,,
 \end{gather}
 where the breaking terms $B_D, B_K$ are, as shown, proportional to $\beta(g)$, with the following proportionality coefficients:
\begin{align}
b_D &= - \del_g G(x_1\ldots x_N)\\
 &=\frac{\mu^{\eps}}{4!} \int \d^d x \,G(x_1\ldots x_N;(\phi^4)_{\rm R}(x)) \,,
 \label{eq:scaleLRI1}\\
b_K &=2 \frac{\mu^{\eps}}{4!} \int \d^d x\, x^\lambda \,G(x_1\ldots x_N;(\phi^4)_{\rm R}(x)) \,.
 \label{eq:scLRI1}
 \end{align}
As shown, $b_D$ can be represented in two equivalent ways. The first way is via the derivative of the correlator with respect to $g$. This is the way leading to the CS equation. The second way is via an integral of $G(x_1\ldots x_N;(\phi^4)_{\rm R}(x))$, which is the correlator of $N$ $\phi$'s with one insertion of the renormalized $\phi^4$ field. On the other hand, $b_K$ has only one representation, of the second type.

It may be surprising that the considered theory satisfies the above Ward identities, given that it does not have a local stress tensor. These identities have been derived in \cite{Paulos:2015jfa} by using the ``Caffarelli-Silvestre trick". This trick consists in formally representing the non-local part of the Lagrangian as a free massless scalar theory living in (in general non-integer) $D=d+2-s$ dimensions. The original $d$-dimensional space where the interaction $\phi^4$ is concentrated is a plane in this $D$-dimensional ambient space. The $D$-dimensional theory being local, it has a stress tensor and enjoys $D$-dimensional scale and conformal invariance. The interaction plane breaks some of the ambient symmetries, but one can still derive Ward identities for the transformations ``tangential to the plane", and restricting those to the plane, one gets the above Ward identities. This is done in detail in \cite{Paulos:2015jfa} and we will not repeat the derivation here.

The second step is to show that the above Ward identities imply conformal invariance.
For this we need to show that the breaking terms $B_D$, $B_K$ are negligible at large distances.
Instead of keeping $\mu_c$ fixed and tending $x_i\to\infty$ we will, equivalently, do the opposite.
 Namely, let us send $g\to g_*$ (for fixed $\mu$). In this limit $\mu_c\to \infty$: the RG flow transitions from UV to IR at very short distances. So at all finite distances $x_i$ the correlators tend to those of LRFP. In this situation, we want to show that the breaking terms approach zero. This will establish that the LRFP correlators satisfy conformal Ward identities.
 
The breaking terms are proportional to $\beta(g)$. Since $\beta(g_*) = 0$, it's enough to show that the coefficients $b_D$, $b_K$ remain finite in the considered limit. All known to us prior literature on the related matters tacitly assumed this fact without any justification.\footnote{See e.g.~\cite{Parisi:1972zy,Brown:1979pq,Braun:2003rp,Vasilev:2004yr}. This literature is dedicated to conformal invariance of IR fixed points of local field theories, but the issue at stake is the same.} 

However, as we explained in \cite{Paulos:2015jfa}, this issue actually deserves discussion. The end result is that $b_D$ and $b_K$ do remain finite, and so we do have conformal invariance. The proof is a bit technical and we relegate it to a separate subsection below. The key idea is to use the fact that $b_D$ has two representations.

\subsubsection{Proof that $b_D$ and $b_K$ remain finite}

That $b_D$ remains finite is actually obvious using the first representation: since correlation functions have a good power series expansion in $g$, the derivative with respect to $g$ at $g=g_*$ is finite, order by order in perturbation theory. This is clearly true when all distances $x_i=O(\mu^{-1})$, so that perturbation theory does not contain large logarithms. For smaller or larger distances we can use CS equation which expresses the correlator in terms of the running coupling. The end result is that
$\del_g G$ for fixed $x_i$, has a finite limit as $g\to g_*$. So $b_D$ is finite.

However, let us examine the finiteness of $b_D$ and $b_K$ for $g\to g_*$ using the second representation (which is the only one available for $b_K$). One realizes that the finiteness is not so obvious from this point of view. Since the field $\phi^4$ is irrelevant at the LRFP, the correlators in the r.h.s.~will have non-integrable powerlike singularities for $x$ near $x_i$. So for $g=g_*$ the integrals in the r.h.s.~are formally powerlike divergent. On the other hand for $g$ somewhat below $g_*$ the integrals are convergent because at $|x-x_i|\lesssim \mu_c^{-1}$ we go to the UV regime where $\phi^4$ is relevant.

A moment's thought shows that the only potentially divergent term in the integrals for $b_D$ and $b_K$ can be due to the $\phi\times \phi^4 \supset \phi$ OPE (other terms in the OPE will give rise to convergent integrals):
\beq
\phi(0) (\phi^4)_{\rm R}(x) \sim C(\mu_c, x) \phi(0)
\eeq
Namely, $b_D$ and $b_K$ will remain finite if and only if
\beq
I=\int d^d x\, C(\mu_c, x) 
\eeq
remains finite in the considered limit.\footnote{Notice that for this argument we use the OPE in the full theory which interpolates between the UV and IR fixed points. Even though this theory has a mass scale, the OPE concept still applies, although it's not as powerful as in conformal theories.} We are using here the fact that the OPE kernel $C$ is rotationally invariant, being the OPE of two scalars, and so when we integrate in the neighborhood of $x_i$ in $b_K$, we will get a contribution proportional to $x_i^\lambda  b_K$, while the term proportional to 
\beq
\int d^d x\, (x-x_i)^\lambda C(\mu_c, x-x_i) 
\eeq
vanishes by rotational invariance.

Now we can complete the argument. We have seen that $b_D$ can only remain finite if $I$ is finite. But we know that $b_D$ remains finite, from the first representation. So $I$ remains finite. But then $b_K$ also remains finite. So we are done.

Let us make a couple more remarks. Notice that we can write, by dimensional analysis
\beq
C(\mu_c, x)= \frac{f(x \mu_c)}{|x|^{[\phi^4]_{\rm IR}}}
\eeq
Then 
\beq
I=\int d^d x \,C(\mu_c, x)=  \mu_c^{[\phi^4]_{\rm IR}-d} \int d^d x\, C(1, x)
\eeq
So the only way $I $ can remain finite is if the integral is actually zero! This is a nontrivial constraint on the OPE kernel, and a curious example of UV/IR conspiracy in quantum field theory. In appendix \ref{sec:OPE} we present an explicit check of this fact at the lowest nontrivial order in $g$. As we find there, the OPE kernel crosses from negative in the UV to positive in the IR, so that the total integral is zero.

If we assume that $I=0$, then we can justify the following prescription for evaluating the integrals in $b_D$ and $b_K$, directly at $g=g_*$. As mentioned, those integrals at $g=g_*$ are formally UV divergent, as a power of UV cutoff. The prescription consists in just dropping this divergence. Our argument that $I=0$ can be seen as a formal justification that this naive prescription is correct.

\subsection{Relative normalization of $\phi$ and $\phi^3$ OPE coefficients}
\label{sec:relope}

\def\D{\Delta}

 As explained in appendix C of \cite{Paulos:2015jfa}, the non-local EOM \eqref{eq:NLEOM}, together with the conformal symmetry, lead to a relation between the OPE coefficients of the operators $\phi$ and $\phi^3$. Let us review this result.
The non-local EOM can be written in position space as
\be
\label{eq:nleomapp}
\int d^d y \frac{1}{|x-y|^{2(d - \Delta_\phi)}} \phi(y) = C\phi^3(x)\,.
\ee
The precise value of $C$ will be unimportant.

 We are interested in the large distance behavior of correlators of $\phi$ and $\phi^3$, which is described by the LRFP. As discussed in the previous section, LRFP is a CFT. The operators $\phi$ and $\phi^3$ are primary operators in this CFT (this is obvious for $\phi$ and also true for $\phi^3$ since there is no operator of which it could be a descendant). 

From the CFT point of view, the overall normalization of operators is not so important. So we introduce operators $\tilde\phi$, $\tilde{\phi^3}$, unit-normalized in the IR. How is their normalization related to that of $\phi$, $\phi^3$? It's convenient to write the 2pt function of $\phi$ in the form:
\be
\label{eq:rho-def}
\vev{\phi(x) \phi(0)} = \frac{1 + \rho(\eps)}{|x|^{2\D_\phi}}\,.
\ee
This and all subsequent equations in this subsection are at large distances so that we are at the LRFP. The coefficient $\rho(\eps)$ is a nontrivial function of $\epsilon$ which is $O(\eps^2)$ as $\epsilon \to 0$. From \reef{eq:Q} we have 
\beq
\label{eq:rho0}
\rho(\eps)=\eps^2 Q/K^2+\ldots\,.
\eeq 
We can now act with the non-local EOM twice on \reef{eq:rho-def} and derive the normalization of $\phi^3$ in the IR. It's important to realize though that the 1 in the numerator does not contribute to the 2pt function of $\phi^3$. This is obvious by looking at which diagrams contribute (see \cite{Paulos:2015jfa}, (3.37) and note \ref{note:subtle}). So we obtain:
\be
\label{eq:phi3phi3app}
\vev{\phi^3(x) \phi^3(0)} = \frac{\rho(\eps) M_2 / C^2}{|x|^{2 \D_{\phi^3}}}\,,
\ee
where $\D_{\phi^3} = d - \Delta_\phi$ consistently with the shadow relation, and
\beq
\label{eq:M2}
M_2 = \frac{\pi^d \G(d/2 - \D_\phi) \G(\D_\phi - d/2)}{\G(d-\D_\phi)\G(\D_\phi)}\,. 
\eeq
Notice that the integral in the non-local EOM is over all distances, while the 2pt functions of $\phi$ used in the calculation holds only at large distances. We are taking a reasonable assumption that the large distance limit commutes with the non-local EOM. This has been shown in \cite{Paulos:2015jfa}
for this 2pt function calculation, but not for the 3pt function calculation performed below. It would be interesting to prove this assumption in full generality.

As discussed in \cite{Paulos:2015jfa}, we expect LRFP to be described by a \emph{unitary} CFT. Indeed, in the considered range of $s$ the field $\phi$ satisfies the unitarity bound in the UV, so we have an RG flow starting at a unitary UV theory perturbed by a hermitian operator.\footnote{Unitarity can also be seen in the lattice long-range Ising model formulation, because the interaction $1/|i-j|^{d+s}$ is reflection positive; see \cite{aizenmanCritical1988}, Eq.~(3.1).} Unitarity of the LRFP implies that we should have nonnegative 2pt functions. Notice also that $M_2<0$ for $0<s<2$. This means that we expect:
\beq
\label{eq:expect}
-1< \rho(\eps) < 0\,.
\eeq
For small $\eps$, this is in agreement with the explicit result \reef{eq:rho0}, since $Q<0$.

Provided \reef{eq:expect} holds, the relation with the unit-normalized fields is as follows:
\be
\label{eq:normal}
\tilde \phi(x) \equiv \frac{1}{\sqrt{1 + \rho(\eps)}} \phi(x)\,, \qquad \qquad \tilde \phi^3(x) \equiv \frac{1}{\sqrt{\rho(\eps) M_2 / C^2}} \phi^3(x)\,.
\ee
The non-local EOM relation between the unit-normalized fields is the same as \reef{eq:nleomapp} but with $C$ replaced by
\beq
\tilde C= \sqrt{\frac{ \rho(\eps) M_2}{1 +  \rho(\eps)}}\,.
\eeq
Not surprisingly, $C$ has canceled out from this relation.

Now let $\calO_1$ and $\calO_2$ be two scalar primary operators of the LRFP CFT (which may or may not coincide with $\tilde\phi$ or $\tilde\phi^3$). By conformal invariance, 3pt functions of $\tilde\phi$ and $\tilde\phi^3$ with these operators take the form:
\begin{gather}
\vev{\tilde\phi(x) \op_1(y) \op_2(z)} = \frac{\lambda_{12\tilde\phi}}{|x-y|^{\D_\phi + \D_1 - \D_2} |x-z|^{\D_\phi - \D_1 + \D_2} |y-z|^{- \D_\phi + \D_1 + \D_2}}\,,
\label{eq:phi12}\\
\vev{\tilde\phi^3(x) \op_1(y) \op_2(z)} = \frac{\lambda_{12 \tilde\phi^3}}{|x-y|^{\D_{\phi^3} + \D_1 - \D_2} |x-z|^{\D_{\phi^3} - \D_1 + \D_2} |y-z|^{- \D_{\phi^3} + \D_1 + \D_2}}\,.
\end{gather}
We would like to use the non-local EOM to relate these two 3pt functions. The diagrams which contribute to the first 3pt function are of two types. Either $\phi$ is connected to an interaction vertex, or it's connected directly to one of the fields inside $\calO_1$ and $\calO_2$. The non-local EOM maps the diagrams of the first type into those of $\phi^3$. Diagrams of the second type depend only on one of the two distances $x-y$, $x-z$. The non-local EOM maps them into local terms proportional to (derivatives of) $\delta(x-y), \delta(x-z)$, irrelevant for the purposes of comparison with the IR behavior.

So, we act with the non-local EOM on the first 3pt function and use Symanzik's star integral formula \cite{Symanzik:1972wj}. We thus obtain an expression of the same functional form as the second 3pt function. This provides a further consistency check of the shadow relation and allows to deduce relative normalization of the 3pt function coefficients \cite{Paulos:2015jfa}:
\be
\label{eq:ratio}
\frac{\lambda_{12\tilde \phi^3}}{\lambda_{12 \tilde \phi}}= \frac{M_3}{\tilde C}=M_3 \sqrt{\frac{1 +  \rho(\eps)}
{ \rho(\eps) M_2}}\,,
\ee
where
\be
\label{eq:M3}
M_3 = \pi^{d/2}\frac{\G(\frac{d - \Delta_\phi + \Delta_{12}}{2}) 
\G(\frac{d - \Delta_\phi - \Delta_{12}}2) 
\G(\Delta_\phi - d/2)}
{\G(d-\Delta_\phi)\G(\frac{\Delta_\phi + \Delta_{12}}2) 
\G(\frac{\Delta_\phi - \Delta_{12}}2)}\,,\quad \Delta_{12} =\Delta_1 - \Delta_2\,.
\ee

Even though this equation depends on $\rho(\eps)$ which is an unknown function of $\eps$ (although it can be computed perturbatively for small $\eps$), notice that it predicts a nontrivial dependence of the ratio on $\Delta_{12}$. In appendix \ref{sec:checksOPEphi4} we perform some checks of this predicted dependence. Taking further ratios we can get relations independent of $\rho(\eps)$. Namely, if we consider two OPEs $\calO_1\times\calO_2$ and $\calO_1'\times\calO_2'$, then 
\be
\label{eq:relative3pt1}
\frac{\lambda_{12 \tilde \phi^3}/\lambda_{12 \tilde \phi}}
{\lambda_{1'2' \tilde \phi^3}/\lambda_{1'2' \tilde \phi}} = \frac{M_3}{M_3'}\,.
\ee

\section{Infrared duality}
\label{sec:IRduality}
\subsection{All-order conjectures about the $\sigma\chi$-flow}
\label{sec:all-order}

We have seen in the previous section how many nontrivial facts about the $\phi^4$-flow can be proved to all orders in the $\eps$-expansion. Here we will give a parallel discussion for the $\sigma\chi$-flow. 
Arguing by analogy, we will motivate  a number of all order results in the $\delta$-expansion. In the next section we will see how it all fits together with the infrared duality.

Compared to the standard perturbation theory of a Lagrangian field theory, whose structural properties are well-understood to all orders,
 conformal perturbation theory is an underdeveloped subject. 
The usual discussion starts from perturbing a CFT by a weakly relevant operator $\calO$ of scaling dimension $d-\delta$, where $\delta\ll 1$. To define perturbation theory, one needs a regulator, and point splitting is a natural choice. However point splitting is awkward to implement at higher orders. Dimensional regularization or analytic regularization in $\delta$ are not viable in general, because the CFT may exist or be tractable only for a fixed spacetime dimension, and because the relevant perturbation usually  exists 
only for a fixed, physical value of $\delta$. %
This is unlike  Lagrangian perturbation theory, where correlation functions can be analytically continued to arbitrary $d$.

Fortunately, the case of the $\sigma\chi$-flow is better than this generic situation, since the dimension of $\chi$ {\it is} a continuously varying parameter -- the gaussian action which governs the dynamics of $\chi$ is defined for any $\Delta_\chi$. So in our case we can consider analytic continuation in $\delta$ as a way to regulate integrals. This provids a potential pathway to an all-order discussion.

We will now discuss how things might plausibly work out in this all-order perturbation theory. The results will be in agreement with the finite order computations that we performed in sections \ref{sec:beta}, \ref{sec:anom} using the point-splitting regulator, and with further subsequent checks. Still, our all-order discussion of the $\sigma\chi$-flow will not reach the level of rigor which was possible for the $\phi^4$-flow. 

The basic object to study are the correlators of $\chi$, defined in perturbation theory by series-expanding the interaction, evaluating the correlation functions in the factorized theory, and integrating using the above-mentioned analytic regulator. Some integrals will produce poles in $\delta$. We conjecture that, to all orders, such poles can be removed by defining the renormalized coupling $g$ related to the bare coupling by the usual relation:
\beq
g_0 =Z_g(g,\delta)\mu^\delta g\,
\eeq
(the function $Z_g$ is of course different from that of the $\phi^4$-flow).
This conjecture seems reasonable because SRFP does not contain any marginal operator. Our computations in sections \ref{sec:beta}, \ref{sec:anom} can be seen as a low-order test. It would be nice to find a full proof.\footnote{We are grateful to David Simmons-Duffin for discussions and for sharing his unpublished notes on all-order conformal perturbation theory.} Assuming the conjecture, we can derive analogues of the all-order $\phi^4$-flow statements by an almost verbatim repetition of the arguments.

A part of the conjecture is that the gamma-function
of $\chi$ is zero. This is motivated in the same way as for $\phi$ in the $\phi^4$ flow. Namely, that poles in $\delta$ correspond to short-distances divergences of the integral for $\delta=0$, the divergences are local, and the action of $\chi$ is non-local, so it can't be renormalized. We then obtain that the anomalous dimension of $\chi$ at the fixed point is identically zero.
This is an all-order generalization of the lowest-order result in section \ref{sec:chi-sigma}.

For future use, notice that if $\chi$ is unit-normalized in the UV, then in the IR we will have
 \begin{equation}
 \label{eq:chinormIR}
  \langle \chi(x) \chi(0) \rangle = \frac{1 + \kappa (\delta)   }{|x|^{2\Delta_\chi}}\, ,
  \end{equation}
 It's clear that $\kappa$ has an expansion in even powers of $g$ so $\kappa(\delta)=O(\delta)$ for small $\delta$. In fact from the lowest order diagram we can easily obtain (using Eq.~(3.14) in \cite{Paulos:2015jfa}):
 \beq
 \kappa = g_*^2 \pi^d \frac{\Gamma
   \left(\frac d2-
   \Delta _{\sigma }\right)
   \Gamma \left(\Delta _{\sigma
   }-\frac{d}{2}\right)}{\Gamma
   \left(\Delta _{\sigma }\right)
   \Gamma \left(d-\Delta _{\sigma
   }\right)}+O(g_*^4)\,.
   \eeq
This is negative, similarly to how $\rho(\eps)$ starts out negative for small $\eps$.

We can argue that the IR fixed point of the $\sigma\chi$-flow should be conformally invariant. Indeed, we can derive the broken conformal Ward identities for the $\sigma\chi$-flow by the same Caffarelli-Silvestre trick. We can then show that these Ward identities imply the conformal invariance in the IR.

 We also have a non-local EOM:
 \begin{equation}
\int d^d y \frac{1}{|x-y|^{2(d - \Delta_\chi)}} \chi(y) = \ C' \sigma (x)\, .
\end{equation}
From this we can see that the shadow relation $\Delta_\sigma+\Delta_\chi = d$ holds at the IR fixed point,
generalizing the lowest-order result in section \ref{sec:chi-sigma}.

Finally, we can repeat verbatim the calculation of section \ref{sec:relope}. Given \reef{eq:chinormIR}, we obtain
  \begin{equation} 
  \label{dualratio}
\frac{   \lambda_{12 \tilde \sigma }} {\lambda_{12 \tilde \chi }   } = \hat M_3 
\sqrt{
\frac{{1 + \kappa(\delta)}}   
{{\kappa(\delta) \hat M_2}}
}\, ,
  \end{equation}
    where $\hat M_2$ and $\hat M_3$ are the ``shadow'' quantities obtained from \reef{eq:M2} and (\ref{eq:M3}) replacing $\Delta_\phi$ with 
  $\Delta_\chi=d-\Delta_\phi$. See appendix \ref{sec:checksLeonardo} for checks of this relation in conformal perturbation theory.

\subsection{Duality interpretation}
\label{sec:dual}

We have seen that the both $\phi^4$-flow and $\sigma\chi$-flow have a conformally invariant IR fixed point. The dimensions of two operators at the fixed point are exactly known (one by non-renormalization, another by the shadow relation):
\beq
\label{eq:phi4res}
[\phi]=(d-s)/2,\quad [\phi^3]=(d+s)/2 ,
\eeq
and
\beq
\label{eq:sigmachi4res}
[\chi]=(d+s)/2,\quad [\sigma]=(d-s)/2.
\eeq
The $\phi^4$-flow relations have been proved to all orders in $\eps=2s-d\ll 1$, near the crossover to mean field.\footnote{\label{note:defect}The fact that the anomalous dimension of $\phi$ is zero can also be seen from the realization of LRFP as a defect CFT via the Caffarelli-Silvestre trick \cite{Paulos:2015jfa}, reviewed in section \ref{sec:confinv}. It follows from the bulk equations of motion together with the assumptions of conformal invariance and bulk-to-defect OPE. See the discussion around Eq.~(4.34) in \cite{Gliozzi:2015qsa}. We are grateful to Pedro Liendo and Marco Meineri for emphasizing this connection to us.}

Under the reasonable assumption of renormalizability, the $\sigma\chi$-flow relations hold to all orders in $\delta=2(s_*-s)\ll 1$, near the crossover to short range.

The most natural interpretation of these results is that there is only one CFTs for each $s$, which describes the fixed points of both flows (infrared duality). The fields $\phi,\phi^3$ for the first flow have to be identified in the IR with $\sigma,\chi$ for the second flow (up to proportionality coefficients). Finally, the above equations for the IR field dimensions are valid \emph{exactly} and not just in perturbation theory. Indeed, if there were nonperturbative corrections, say, to the first set of equations, they would presumably become largest near the short-range crossover, but this is where the second set of equations becomes accurate and shows that there are no corrections.

\begin{figure}[h]
\centering
  \includegraphics[width=.6\textwidth]{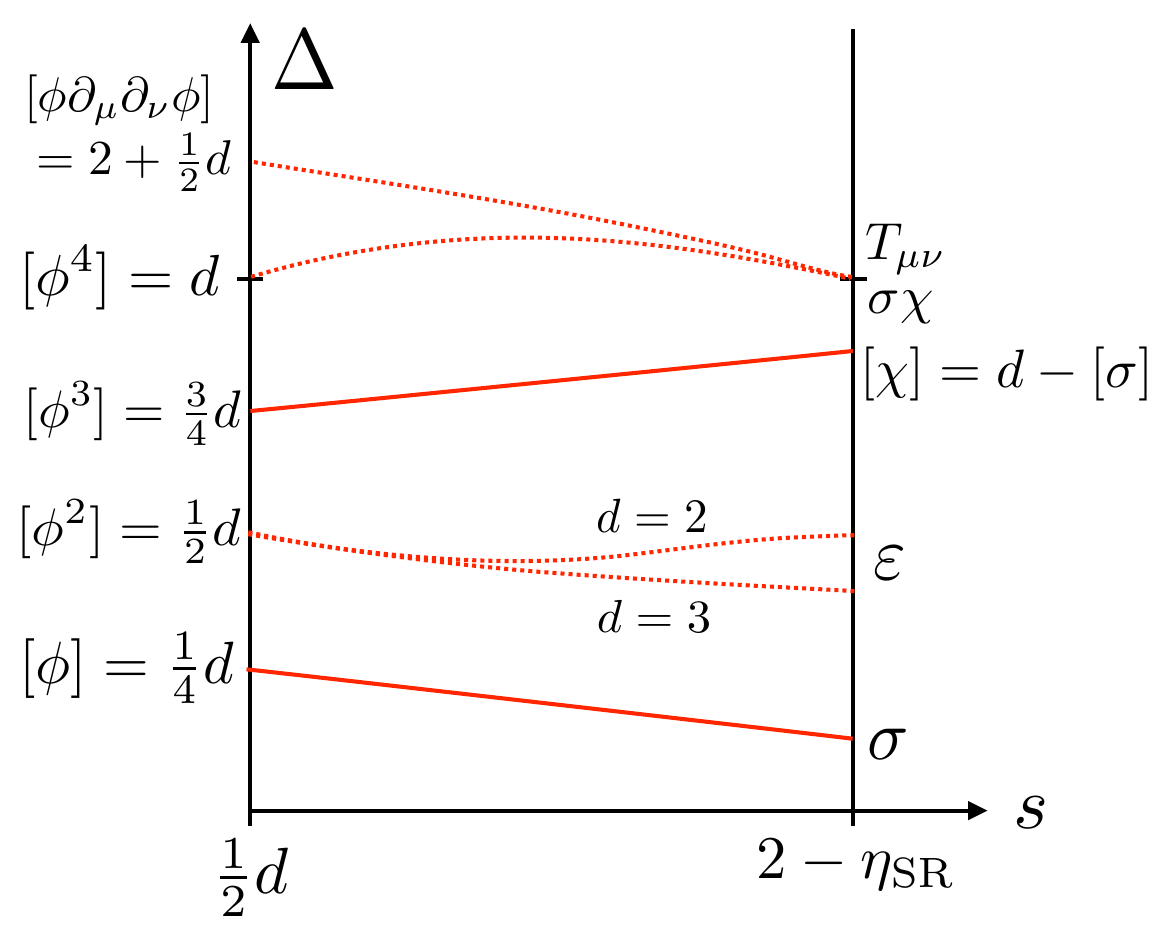}
  \caption{The dependence of dimensions of several important operators on $s$.}
  \label{fig:spectrum}
 \end{figure}

See Fig.~\ref{fig:spectrum} for the predicted dependence of the most important LRFP operator dimensions on $s$ between the mean-field and the short-range crossovers. The solid lines joining $\phi$ to $\sigma$ and $\chi$ to $\phi^3$ are straight lines. The other lines are known only approximately in the $\eps$ and $\delta$ expansion around the crossovers. The shown shape of the lines is the simplest consistent with these asymptotics. The line joining $\phi^2$ to $\varepsilon$ deserves a comment. We have $[\phi^2]=(d-\eps)/2+\gamma=d-\eps/6+O(\eps^2)$ near the mean-field crossover,
so that the line starts going linearly down. In $d=3$ it joins to $\Delta_\vareps\approx 1.41<d/2$ with the negative first derivative, see \reef{eq:gammaeps}. However, in $d=2$ it rises back up to $\Delta_\vareps=1=d/2$ and has a zero first derivative at the crossover, see \reef{eq:gammaepsd=2}.

As a further check of the duality, we will show that the relations (\ref{eq:ratio}) and (\ref{dualratio}) for the OPE coefficients can be made compatible with each other. We must have
\begin{equation}
\label{eq:musthave}
\frac{\lambda_{12 \tilde \phi^3}}{\lambda_{12 \tilde \phi}}  \equiv \frac{   \lambda_{12 \tilde \chi }} {\lambda_{12 \tilde \sigma }   } \, .
\end{equation}
It's not trivial that the two sides can agree, because the dependence on $\Delta_{1,2}$
must match. Fortunately it does, thanks to the following identity,
\beq
M_3  \hat M_3 = M_2 = \hat M_2\,.
\eeq
 Equation~\reef{eq:musthave}  holds provided that $\rho$ and $\kappa$ obey, at the same value of $s$, the relation:
\begin{equation}
\frac{\kappa(\delta) }{1 + \kappa(\delta)} =  \frac{1 + \rho(\epsilon)}{\rho(\epsilon)}\,.
\end{equation}
This leads to nontrivial predictions for the behavior of the two functions near the mean-field and short-range crossovers. Since as we have seen $\kappa(\delta)=O(\delta)$ for small $\delta$, we conclude that the normalization of the 2pt function of $\phi$ must vanish linearly in $s$ close to the short-range crossover:
\beq
1+ \rho(\eps)=O(s_*-s).
\label{eq:normvan}
\eeq
This was previously discussed as a necessary condition for a smooth crossover in \cite{Paulos:2015jfa}, Appendix C.\footnote{As the very last sentence 
of \cite{Paulos:2015jfa} shows, at the time this vanishing seemed a bit paradoxical. Now we feel totally comfortable with this conclusion, as it is an unambiguous prediction of our picture.} The same vanishing of the 2pt function normalization has been previously argued in \cite{Parisi}, via a completely different argument. {See also appendix \ref{sec:log} for a discussion of the $\langle\phi\phi\rangle$ correlator precisely at the crossover, where it exhibits a logarithmic suppression with respect to the naive scaling. We derive the suppression via RG, but it can also be thought of as a remnant of the vanishing of the normalization (\ref{eq:normvan}) \cite{Parisi}.}

Analogously, we must have 
\beq
1 + \kappa(\delta) = O((2s-d)^2)
\eeq
when approaching the crossover to the mean-field regime. In this case the vanishing is expected to be quadratic since $\rho(\eps)=O(\eps^2)$.

In summary, we have accumulated strong evidence for a novel IR duality: the $\phi^4$-flow and the $\sigma \chi$-flow end in the same IR fixed point.
Unlike all previously
studied examples of IR dualities, our theories lack a local stress tensor. Non-locality, which is an essential feature of our construction, comes with a surprising bonus: remarkable computation power! One of its immediate consequences is the non-renormalization theorem 
for $\Delta_\phi$ and  $\Delta_\chi$. The non-local equations of motions were then used to show  that $\Delta_{\phi^3}$ and of $\Delta_\sigma$ obey shadow relations at the IR fixed point, and that
OPE coefficients involving the shadow pairs must come in precise ratios. In
 the paradigmatic examples of IR dualities  (such as 4d Seiberg duality and 3d mirror symmetry), it is supersymmetry that gives analytic control. Curiously, 
 we were able to achieve significant analytic control in our non-supersymmetric setting,  thanks precisely to the non-local nature of the problem.


\section{Discussion}
\label{sec:conclusions}
 
 In this paper we have proposed and studied a new compelling picture for the long-range to short-range crossover. 
Prior to our work, the understanding of this crossover was incomplete at best.
Some  of its qualitative features   -- in particular its continuous nature -- had been anticipated, but doubts remained, 
as evidenced by some recent controversies in the literature  (see appendix \ref{sec:prior}). Other important features of the crossover were completely missed, in particular 
the fact that the crossover happens not to the SRFP, but to the SRFP plus a decoupled gaussian field.

Crucially, our new qualitative picture   allowed us to advance greatly the quantitative side of the story, hitherto non-existent. We obtained a number of predictions for the critical exponents near the crossover, which in principle can be confirmed by Monte Carlo simulations and, perhaps, experiments. Hopefully this 
would 
convince the remaining skeptics that the crossover is continuous. 

The infrared duality between the $\phi^4$ and $\sigma\chi$-flows is essential 
to our picture.  All our findings support this idea. Notably, we have seen that both flows contain in the IR a pair of operators $(\calO_1,\calO_2)$ satisfying the shadow relation $\Delta_1+\Delta_2=d$. We argued that these relations are true to all orders in perturbation theory, and in view of the duality the simplest assumption is that they are also valid non-perturbatively.  The shadow relation and the related results about the normalization of OPE coefficients (see section \ref{sec:dual})
will prove useful in the analysis of the LRFP using the conformal bootstrap. Work in this direction is ongoing \cite{Behan}.\footnote{The 3d LRFP is also expected to have a $\mathbb{Z}_2$ line defect operator, analogous to the SRFP line defect \cite{Billo:2013jda, Gaiotto:2013nva} and continuously connected to it. This may explain why some ongoing bootstrap studies \cite{Mazac} do not succeed in isolating the 3d SRFP line defect. We thank Dalimil Maz\'a\v{c} for this remark.}

While in this paper we have focused on the long-range \emph{Ising} model, it's clear that most of the learned lessons are quite general. For example, the extension to the $O(N)$ case is straightforward. Still more generally, our $\sigma\chi$-flow construction can be used 
with any CFT in place of the SRFP. Just pick a scalar CFT operator, call it $\sigma$ again, of dimension $\Delta$, and couple it to a non-local gaussian field $\chi$ of dimension $d-\Delta-\delta$, $\delta\ll1$. One then needs to compute the quantum correction to the beta-function. 
Naively, there is a 50\% chance that the quantum correction has the right sign to yield a stable IR fixed point. We will then obtain a continuous family (parametrized by $\delta$) of non-local conformally invariant theories which are deformations of the original local CFT. It will be unitary if the original CFT was unitary and if $\chi$ is above the unitarity bound. The generic prediction of this construction is the non-renormalization of $\chi$ and the IR shadow relation $\Delta_{\sigma,\rm IR}+\Delta_\chi=d$.

This demonstrates that, while we expect local CFTs to be generically isolated, non-local conformal theories can easily form continuous families. While this observation by itself is not new, the above general construction seems new. Another known way to construct such continuous families is to put a UV-complete massive theory in a fixed AdS background. Varying masses and couplings of the bulk theory, we obtain a continuous family of boundary theories which have conformal invariance but no local stress tensor (since the metric is non-dynamical). See e.g.~\cite{Paulos:2016fap}.

We would like to finish with a brief discussion of the $d=1$ case, which has so far been excluded from our considerations (except in section \ref{sec:standard}). The $d=1$ short-range Ising model does not have a phase transition, and so the physics of long-range to short-range crossover is bound to be very different from $d=2,3$. The only scale-invariant phase of the $d=1$ short-range Ising model occurs at zero temperature, where all correlation functions are constant. This corresponds to the commonly assigned exact critical exponent $\eta=1$, see e.g.~\cite{Holovatch1993} (recall that $\Delta_\phi=-1/2$ is the scalar field engineering dimension in $d=1$).
Applying naively the general $d$-dimensional formula \reef{s*} for the crossover location, we expect it to happen at $s=2-\eta=1$. This matches nicely with what is known about the long-range Ising model in $d=1$. First of all, since the work of Dyson \cite{dyson1969} it is rigorously known that model has a phase transition for $0<s<1$. This transition is continuous in this range, as is also rigorously known (\cite{Hugo}, Corollary 1.5). The transition disappears for $s>1$, which is where we expect the short-range phase. 

The borderline case $s=1$ is special: the phase transition exists, but it's discontinuous, in the sense that the magnetization has a nonzero limit for $\beta\to\beta_c^+$, as was argued by Thouless \cite{Thouless} and later proved rigorously \cite{ACCN}. This phase transition is topological, driven by dissociation of kink-antikink pairs \cite{Thouless,AndersonYuval,Scalapino}.\footnote{The work of Thouless \cite{Thouless} foreshadowed later work by Berezinski and by Kosterlitz and Thouless on the BKT phase transition in two-dimensional $O(2)$ models, driven by dissociation of vortex-antivortex pairs.} For $s=1$ ($1/|x|^2$ spin-spin interaction), kinks interact logarithmically. As temperature is raised, defect operators representing kinks become relevant, kinks proliferate, and the model disorders.  So, the theory at $s=1$ and $\beta=\beta_c$ has marginally relevant operators (kinks). This suggests that the LRFP at $s$ slightly below 1 should have a weakly coupled description. We hope to return to the problem of finding such a description in the future.
 
\section*{Acknowledgements}

We thank Michael Aizenman, Barry McCoy, Pedro Liendo, Dalimil Maz\'a\v{c}, Marco Meineri, Giorgio Parisi and Gordon Slade for useful comments and discussions. We are grateful to the Galileo Galilei Institute of Theoretical Physics where this work was initiated, and to the Princeton Center for Theoretical Science where it was first presented. SR is supported by Mitsubishi Heavy Industries as an ENS-MHI Chair holder. SR and BZ are supported by the National Centre of Competence in Research SwissMAP funded by the Swiss National Science Foundation. LR and SR are supported by the Simons Foundation grants 397411 and 488655 (Simons collaboration on the Non-perturbative bootstrap). CB is supported by the Natural Sciences and Engineering Research Council of Canada.

\appendix
\section{Off-critical behavior of the long-range model}
\label{sec:dis}

This appendix concerns the long-range model off criticality. Just like in the case of the ordinary, nearest-neighbor, Ising model, the critical point of the long-range model separates the disordered phase from the ordered phase, the latter characterized by a nonzero value of the order parameter $\langle\phi\rangle$, the mean value of $\phi$.
 
The disordered phase is described by the action \reef{standardflow} with a positive quadratic term coefficient $t>0$, while the quartic term is unimportant (see below). In momentum space we find\footnote{For the long-range model with $s=2$ this equation needs to be modified as $|p|^s\to p^2\log|p|$. The important point is that this function is non-analytic at $p=0$, and the arguments below leading to the power law decay go through.}
\beq
\label{eq:disord}
S=\int \frac{d^dp}{(2\pi)^d}[C|p|^s+t]\phi(p)\phi(-p),
\eeq
where $C=-w_{d+s}$ and $w_A$ is the constant in the basic integral
\beq
\int d^dx\, \frac{e^{ipx}}{|x|^A}=\frac{w_A}{|p|^{d-A}},\quad w_A=(2\pi)^{d/2} 2^{d/2-A}\frac{\Gamma((d-A)/2)}{\Gamma(A/2)}\,.
\eeq
Notice that $C>0$ in our primary interest range $0<s<2$ (see note \ref{note:0s2}). We will rescale fields and momenta and set $C$ and $t$ to one in the rest of this section.

Consider now the 2pt function of $\phi$ in position space:
\beq
\label{eq:G00}
G(x)=\int \frac{d^dp}{(2\pi)^d}\, \frac{e^{ipx}}{|p|^\mysigma + 1}\,.
\eeq
Since the propagator is nonanalytic at $p=0$, $G(x)$ will have power law decay. At short distances $x\ll1$ the integral is dominated by $p\gg1$ where we can neglect 1 in the denominator and obtain
\beq
G(x)\sim \frac{1}{(2\pi)^d}\frac{w_s}{|x|^{d-s}}\qquad(x\ll 1)\,,
\eeq
consistently with $[\phi]_{\rm UV}=(d-s)/2$.
On the other hand, at long distances the integral is dominated by $p\ll 1$ where we can expand the propagator as
 \beq
\frac1{|p|^\mysigma + 1}=1-|p|^\mysigma+\ldots
\eeq
The leading constant term gives a purely local $\delta(x)$ contribution to $G(x)$, while the leading long-distance contribution comes from the next term:
\beq
\label{eq:2pt-para}
G(x)\sim -\frac{1}{(2\pi)^d}\frac{w_{-s}}{|x|^{d+s}}\qquad(x\gg1)\,.
\eeq
This gives the IR dimension of $\phi$ in the disordered phase, $[\phi]_{\rm dis}=(d+s)/2$.\footnote{This result has been established rigorously for the long-range Ising model on the lattice, see section \ref{sec:rigor}.} Notice that the $\phi^2$ interaction is irrelevant for this value of the $\phi$ dimension, showing the full IR stability of the disordered fixed point.
Further interactions like $\phi^4$ are even more irrelevant, justifying their neglect in the above discussion.

{{It has to be pointed out that while action \reef{eq:disord} captures qualitatively the change of the $\phi$ dimension along the flow to the disordered phase, it cannot be quantitatively correct at intermediate distances of the order of the correlation length, where the neglected quartic term is important. On the other hand action \reef{eq:disord} is appropriate to describe the physics at distances much larger than the correlation length.}}

For the ordered phase the only role of the quartic interaction is to stabilize $\langle\phi\rangle$. Fluctuations around the mean value can then be described by a quadratic action with a positive mass term. The IR scaling dimension of these fluctuations is thus the same as in the disordered phase. 

{The above power law behavior can be contrasted with the 2pt function of the nearest-neighbor model in the disordered phase, which can be obtained by setting $\mysigma=2$ in Eq.~\reef{eq:G00}. In the nearest-neighbor case, the propagator is an {\it analytic} function of momenta with poles in the complex plane, which leads to the {\it exponential} decay of the 2pt function at long distances.} 

\section{The $\langle \phi \phi\rangle$ 2pt function at the crossover}
\label{sec:log}

Consider the LRI precisely at the crossover, $s=s_*$. Since the IR theory contains a marginally irrelevant operator, we expect that the 2pt functions will behave like a power law with logarithmic corrections. To determine the power of log, we go to the dual picture and study the $\langle\sigma\sigma\rangle$ correlator in presence of the $\sigma\chi$ perturbation. We keep $\delta=0$ (crossover). 
The CS equation predicts that the correlator depends on the distance as
\beq
\langle\sigma(r)\sigma(0)\rangle= r^{-2\Delta_\sigma} c(r)\,,
\eeq
where 
\beq
c(r)= \hat c(\bar g) \exp\left \{-2\int_1^r d\log r'\, \gamma(\bar g(r',g_0))\right \}=\hat c(\bar g) \exp\left \{2\int_{g_0}^{\bar g} dg\,\frac{\gamma (g)}{\beta(g)}\right\}\,.
\label{eq:c}
\eeq
Here $g_0$ is the coupling at $r=1$ where we start the RG flow, and $\bar g(r,g_0)$ is the running coupling.

The function $\hat c(\bar g)= 1+O(\bar g^2)$ is obtained by matching with perturbation theory; we will only need the leading term.
The leading beta-function is $\beta(g)=-(A/6) g^3$ ($A<0$), and the gamma-function of $\sigma$: $\gamma_\sigma(g)=-(B/2) g^2$, $B=A/3$.
This implies
\beq
\label{B.3}
c(r)\approx \exp\left\{2\int_{}^{\bar g} dg/g\right\} = \bar g(r)^2\propto 1/(\log r)\,,
\eeq
where we used that, by the beta-function equation, the coupling runs to zero as $\bar{g}\propto 1/\sqrt{\log r}$ at large distances.

This logarithmic suppression of the 2pt function at the crossover has been also argued in \cite{Parisi}. It is natural in view of the fact that the IR normalization of the $\langle\phi\phi\rangle$ 2pt function vanishes linearly when approaching the crossover point from below (section \ref{sec:dual}).

The case of small $\delta>0$ can be considered similarly, giving:
\begin{equation}
c(r)\sim \frac{\delta}{e^{2\delta \log(r/r_0)}-1}\,.
\end{equation}
The $\delta\to 0$ limit agrees with \reef{B.3}. This crossover behavior was conjectured in  \cite{Parisi}.

\section{Selected prior work on the long-range Ising model}
\label{sec:prior}

\subsection{Physics}
\label{sec:priorphys}
The study of the long-range Ising model has started in earnest in \cite{Fisher:1972zz}, where also the effective description based on the $\phi^4$-flow has been proposed. That reference has erroneously put the crossover to short range at $s=2$. This was corrected by \cite{Sak,Sak77}, leading to what we called the ``standard picture", which has since been supported by theoretical studies \cite{Honkonen:1988fq,Honkonen:1990mr} and by lattice Monte Carlo simulations \cite{Blote}.

More recently some debate restarted about the nature of the long-range to short-range crossover.
Lattice Monte Carlo simulations in \cite{Picco:2012ak} observed deviations from the standard picture near the crossover (see also \cite{Blanchard:2012xv}) However, Ref.~\cite{Picco:2012ak} may have underestimated systematic errors due to possible logarithmic corrections to scaling near the crossover \cite{Angelini}. From our perspective these logarithmic corrections are associated with the operator $\sigma\chi$, marginally irrelevant at the crossover (see section \ref{sec:existence}).

Ref.~\cite{Defenu:2014bea} analyzed the problem using the functional renormalization group and also found support for the standard picture. 

As a side remark, some of this recent literature likes to phrase the conclusions in terms of the so called ``effective dimension" $D_{\rm eff}(s)$ such that the LRFP in $d$ dimensions is supposed to have the same critical exponents as the SRFP in $D_{\rm  eff}$ dimensions. We would like to use this occasion to stress that this ``effective dimension" is clearly not a fundamental notion, bound to work only for a few exponents and only in low orders of perturbation theory. 

Many of the above-mentioned papers also considered the $O(N)$ generalization of the long-range Ising model. Recently, Ref.~\cite{Parisi} analyzed the crossover in the large $N$ approximation. They argued that the IR normalization of the 2pt function of $\phi$ vanishes as $s\to s_*$. While we do not fully understand the details of their argument, the conclusion agrees with our picture, as discussed in section \ref{sec:dual}. See also \cite{Gubser:2017vgc} for a recent discussion of the large N limit in this model.

\subsection{Rigorous results}
\label{sec:rigor}

The $\phi^4$-flow has been studied via rigorous renormalization group analysis in $d=1,2,3$ in \cite{Brydges:2002wq,Abdesselam:2006qg,Mitter-talk,Slade}. These works show that an infrared fixed point exists nonperturbatively at least for sufficiently small $\eps>0$. For some critical exponents, dependence on $\eps$ in this region of small $\eps$ has also been rigorously investigated. Ref.~\cite{Mitter-talk} announced a proof that $\phi$ does not acquire anomalous dimension. Ref.~\cite{Slade} showed that the susceptibility and the specific heat critical exponents take, at leading order in $\eps$, values expected from the $\eps$-expansion predictions for $\gamma_\phi$ and $\gamma_{\phi^2}$ (this reference considers the general $O(n)$ case, including $n=0$ corresponding to self-avoiding walks).
 
Long-range Ising model can also be studied directly from the lattice Hamiltonian, without relying on the renormalization group. It is known that the model has a phase transition separating a low-$\beta$ phase with vanishing magnetization from a high-$\beta$ phase where the magnetization is nonzero. Moreover the transition is continuous, in the sense that magnetization vanishes as $\beta_c$ is approached from above. The above statements have been proved rigorously for $d=1$, $0<s<1$ and for $d=2,3$, $0<s<2$, which includes the range $0<s<2-\eta_{\rm SR}$ we are interested in (\cite{Hugo}, section 1.4). Incidentally, the same paper also proved for the first time continuity of the phase transition in the \emph{short-range} Ising model for $d=3$.

Another rigorous result worth mentioning is that the spin-spin correlator of the long-range Ising model decays at $\beta<\beta_c$ with the exponent $d+s$, as in our Eq.~\reef{eq:2pt-para}. 
See \cite{aizenmanCritical1988}, Eq.~(2.8). Many other rigorous results about the long-range Ising model on the lattice are reviewed in that paper. 

\section{Integrals for $\gamma_\vareps$ and $\gamma_T$} 
\label{sec:integral}

\begin{figure}[h]
\centering
  \includegraphics[width=.3\textwidth]{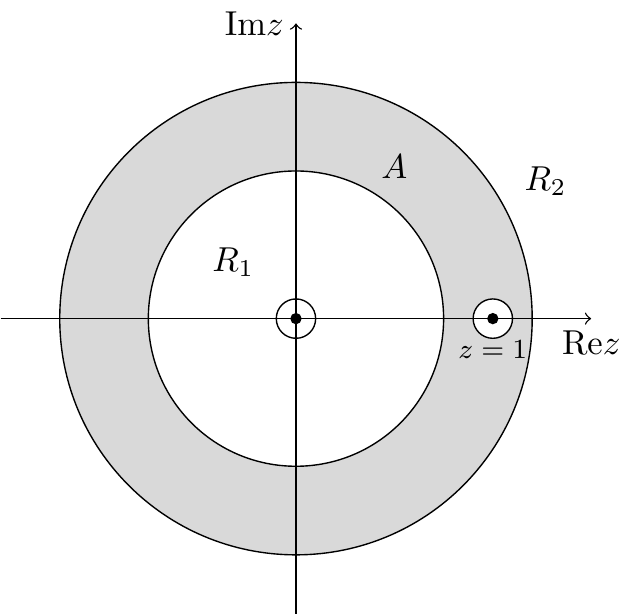}
  \caption{The three integration regions \reef{eq:threeregions}.}
  \label{fig:sub1}
 \end{figure}

In the computation of $\gamma_\varepsilon$ in $d=2$ we encountered a vanishing integral
\beq
\int_\bC d^2 z\, f(z,\zb)=\int_\bC d^2 z \frac{1}{|1-z|^{4}}\left(\frac{|1+z|^2}{4 |z|}-1\right)=0\,, \label{eq:intapp}
\eeq
Let us give an analytic proof of this fact.
It is important to remember that this integral is not absolutely convergent, and needs to be computed with circular cutoffs around $0$, $1$ and infinity. We divide the complex plane into three regions (see Fig.~\ref{fig:sub1}): 
\beq
\begin{aligned}
R_1&=\left\lbrace z:\eps<|z|<1-\delta\right\rbrace\,,\\
A&=\left\lbrace z: 1-\delta<|z|<1+\delta, |z-1|>\eps \right\rbrace\,,\\
R_2&=\left\lbrace z:1+\delta<|z|<\eps^{-1} \right\rbrace\,.\label{eq:threeregions}
\end{aligned}
\eeq

We need to compute the integral for small but finite values of $\eps$ and 
then take $\eps\to0$ limit. The quantity $\delta$ is introduced for convenience. In principle the sum of three integrals does not depend on it, but we will see that all three integrals will simplify for $\delta\ll 1$, so we will take a limit $\delta\to0$ (after $\eps\to0$). It will turn out that the contribution of the region $A$ approaches a nonzero constant for $\delta \to 0$. It's easy to forget about this contribution and get a wrong answer.

\begin{figure}[h]
  \centering
  \includegraphics[width=.2\textwidth]{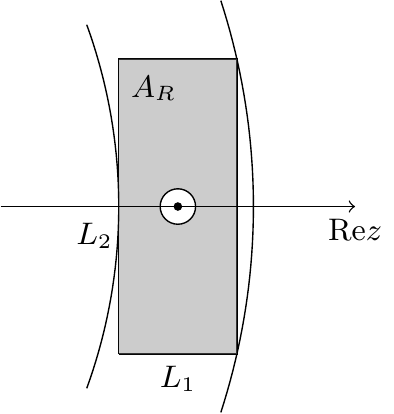}
  \caption{Deformation of the region $A$, which yields the same result in the $\delta \to 0$ limit.}
  \label{fig:sub2}
\end{figure}

The integrals over $R_1$ and $R_2$ can be computed by writing 
\beq
\int d^2z f(z,\bar z)= \int r dr d\theta f(r e^{i\theta},r e^{i\theta}) = \int r\,dr\oint \frac{d\rho}{i\rho} f(r \rho, r/\rho)
\eeq
and doing the $\rho$ integrals by residues. This way one obtains:
\beq
\begin{aligned}
\lim_{\eps \to 0}\int_{R_1} d^2 z\,f(z,\zb)=\frac{\pi}{8}+O(\delta)\,,\\
\lim_{\eps \to 0}\int_{R_2}d^2 z\,f(z,\zb)=\frac{\pi}{8}+O(\delta)\,.\label{eq:r1r2}
\end{aligned}
\eeq
We are left with computing the integral over the region $A$. When the limit $\delta \to 0$ is taken, and the annulus shrinks, the integral will give a non zero contribution because of the singularity at $z=1$. We can restrict the integration region $A$ to a rectangle around $z=1$, as the regions where the integrand is not singular yield a zero contribution in the $\delta \to 0$ limit. We consider therefore the region in Fig.~\ref{fig:sub2}.

We expand the integrand around $z=1$ and keep only the divergent terms, since only they contribute in the $\delta \to 0$ limit. Doing the shift $z\to 1+ z$ and defining the region shown in Fig.~\ref{fig:sub2}, $A_R=\left\lbrace z:-\frac{L_1}{2}<\text{Re}\, z<\frac{L_1}{2}, -\frac{L_2}{2}<\text{Im}\, z <\frac{L_2}{2},|z|>\eps\right\rbrace$, we have
\beq
\lim_{\delta \to 0} \lim_{\eps \to 0} \int_A d^2 z\,f(z,\zb)=\lim_{\delta \to 0} \lim_{\eps \to 0} \int_{A_R}d^2z \left(\frac{1}{8z^2}+\frac{1}{8\zb^2} \right) \,.\label{eq:intannulus}
\eeq
It's straightforward to carry out the integration of the r.h.s., and one obtains
\beq
\int_{A_R}d^2z \left(\frac{1}{8z^2}+\frac{1}{8\zb^2}\right)=\frac{1}{4}\left(\pi -4\tan^{-1}\frac{L_2}{L_1} \right)\,.
\eeq
The result does not depend on the cutoff $\eps$ once we carry out the angular integration. In order to take the $\delta \to 0$ limit we need to understand how $L_1$ and $L_2$ scale with $\delta$. We see that $L_1 \sim \delta$ and $L_2 \sim \sqrt{\delta}$, therefore $\tan^{-1} L_2/L_1 \to \pi/2$ when $\delta \to 0$. Therefore, using \reef{eq:intannulus},
\beq
\lim_{\delta \to 0} \lim_{\eps \to 0} \int_A d^2z\,f(z,\zb) = -\frac{\pi}{4}\,.
\eeq
Combining this result with \reef{eq:r1r2},
\beq
\int_\bC d^2 z\, f(z,\zb)=\lim_{\delta \to 0} \lim_{\eps \to 0} \left(\int_{R_1}+\int_{A}+\int_{R_2} \right)f(z,\zb)\,d^2z=0\,.
\eeq

The same line of reasoning gives the result \reef{eq:TInt} for the anomalous dimension of the stress tensor. Given the integral
\beq
\int_\bC d^2 z\, g(z,\zb)= \int_\bC d^2 z \frac{1}{(1-\bar{z})^{2}}\frac{(z^{2}+30z+1)}{z^{2}}\,,
\eeq
the contributions of the three regions \reef{eq:threeregions}, in the limit $\eps \to 0$ and $\delta \to 0$, are
\beq
\begin{gathered}
\int_{R_1} d^2z\, g(z,\zb)=\int_{R_2} d^2z\, g(z,\zb) = \pi\,,\qquad \int_{A} d^2z\, g(z,\zb)=-32\pi\,.
\end{gathered}
\eeq
This gives us the result \reef{eq:TInt}.

\section{$\phi\times \phi^4$ OPE}
\label{sec:OPE}

Consider the OPE
\beq
\phi(x) \times \calO(0) \sim C(x) \phi(0),\quad \calO=\frac 1{4!}(\phi^4)_{\rm R}\,.
\eeq
Consider the integral $I$ of the kernel $C(x)$ over all space. Notice that the integral is convergent, since $\calO$ is relevant in the UV and irrelevant in the IR. In section \ref{sec:confinv} we gave a general consistency argument that this integral is exactly zero. This fact may appear strange at the first sight, as it requires a delicate cancellation between the UV and IR regions. In this appendix we will provide an explicit perturbative check, showing
that $I$ vanishes to the lowest nontrivial order in $\eps$. 

Unlike in Section \ref{sec:confinv}, here we will not need to take any particular limit $g\to g_*$. We will consider the flow at fixed $g$ and $\mu$,
and will construct $C(x)$ at all distances, all the way from the UV fixed point to IR fixed point. So roughly $\mu$ in this section is $\mu_c$ in section \ref{sec:confinv}.

\begin{figure}[h]
\centering
\includegraphics[width=0.8\textwidth]{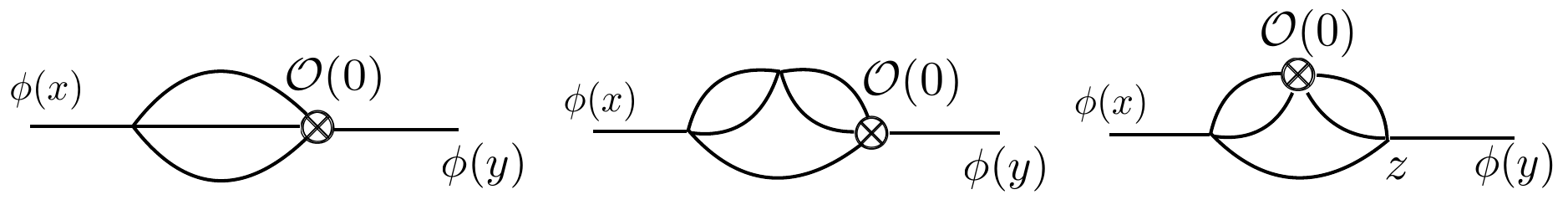}
\caption{Diagrams contributing to the OPE computation.}
\label{fig-OPE}
\end{figure}

To compute the OPE kernel we consider the correlation function $\langle \phi(x) \calO(0)\phi(y)\rangle$. Up to order $g^2$, the leading contribution in the limit $x\ll y$ comes from three diagrams in Fig.~\ref{fig-OPE}. There are other order $g$ and $g^2$ diagrams, but they are subleading for $x\ll y$ and interpreted in terms of other OPEs. For example the mirror image of the first diagram contributes to the OPE $
\phi(x) \times \calO(0) \supset \phi^3(0)$, less singular for $x\to0$.

The first two diagrams are easily computed using repeatedly Eq.~(3.15) in \cite{Paulos:2015jfa}. The third diagram is hard to compute for general $y$, but in the limit $y\ll x$ we can neglect variation of the $\langle \phi(y) \phi(z)\rangle$ propagator with $z$, since the relevant integration region of the ``hard" subdiagram will have $z\sim x$. Replacing
\beq
\langle \phi(y) \phi(z)\rangle \approx \langle \phi(y) \phi(0)\rangle\,,
\eeq
this diagram is then also easily computed in the same way as the other two. 
The bare OPE kernel up to the second order is then given by
\beq
\label{eq:ope-bare}
C_{\rm bare}(x) = g_0 D_1/|x|^{d-2\eps}+g_0^2 (D_2+D_3)/|x|^{d-3\eps}\,,
\eeq
where the values $D_i$ of the three diagrams, taking into account the signs and the symmetry factors are given by (see \cite{Paulos:2015jfa} for the definition of $w_A$) 
\begin{align}
  D_1&=-\frac{1}{3!}\frac{w_\frac{d-\epsilon}{2} w_{3\frac{d-\epsilon}{2}}}{w_{d-2\epsilon}}\approx
  -Y\eps,\quad Y = \frac{1}{3!}\, \pi^{d/2}\frac{\Gamma(-d/4)\Gamma(d/2)}{\Gamma(3d/4)}\,,\nn\\
  D_2&=\frac{1}{4}  \frac{w_{d-\epsilon}^2 w_\frac{d-\epsilon}{2} w_{\frac{3}{2}d - \frac{5}{2}\epsilon}}{w_{d-2\epsilon} w_{d-3\epsilon}}
  \approx 
  3YK  \,,\quad K=(3/2)\Sd\,,
\nn\\
  D_3&=\frac{1}{4}  \frac{w_{d-\epsilon} w^2_\frac{d-\epsilon}{2} w_{\frac{3}{2}d - \frac{5}{2}\epsilon}}{w_{\frac{d-3\epsilon}{2}} w_{d-3\epsilon}}
  \approx 
  \frac{3}{2} YK   \,.
  \label{eq:R1R2}
  \end{align}
To compute $I$ at the lowest nontrivial order we will need only the shown approximate expression for small $\eps$. Notice that although $D_2$ and $D_3$ are suppressed by an extra power of $g_0$ in \reef{eq:ope-bare}, they will end up contributing at the same order in $\eps$ as $D_1$ since $D_1=O(\eps)$. We factored out the common proportionality coefficient $Y$ for further convenience, and reduced $D_{2,3}$ to the second order beta-function coefficient $K$. Notice that $D_1$ and $D_2$ are, up to different combinatorial factors, the same as the diagrams encountered in \cite{Paulos:2015jfa} in the computation of the $\langle \phi\phi^3\rangle$ correlator.

To get the OPE kernel at all distances, we must perform the renormalization group improvement. So first of all we replace in \reef{eq:ope-bare} the coupling $g_0$ by the renormalized coupling $g$ via Eq.~\reef{eq:g0}. The function $Z_g$ is given, at lowest orders, by (see \reef{eq:beta-stan})
\beq
Z_g=1+Kg\eps^{-1}+\ldots.
\eeq
We also have to replace $\calO$ by $\calO_{\rm R}$, related by (see \cite{Paulos:2015jfa}, (2.17))
\beq
\label{eq:Zphi4}
\calO = Z_{\phi^4} \calO_{\rm R},\quad Z_{\phi^4} = 1-K_4 g\eps^{-1} +\ldots, \quad K_4 = 2K\,.
\eeq
The OPE kernel for the renormalized operator and in terms of the renormalized coupling is then given by (cf.~(3.18) in \cite{Paulos:2015jfa}) 
\begin{gather}
\label{eq:chat}
C(x,\mu,g)= \frac{1}{|x|^{d-\eps}} c(r,g)\,,\quad r=\mu x\,,
\\
\label{eq:cgs}
c(r,g)=g r^\eps D_1 +g^2 ( D_1(K+K_4)r^{\eps}\eps^{-1}+ (D_2+D_3)r^{2\eps}) +O(g^3)\,.
\end{gather}
Eq.~\reef{eq:cgs} as written is only useful for $x = O(\mu^{-1})$. To get the OPE kernel at larger or smaller distances we have to solve the corresponding CS equation (see \cite{Collins:1984xc}, Section 10.5)  
\beq
(r\,\del_r +\beta(g)\del_g +\gamma_{\phi^4}(g)) c(r,g) = 0\,.
\eeq
As is well known the solution can be written in terms of the running coupling $\bar g=\bar g(r,g)$ from Eq.~\reef{eq:running},
\beq
\label{eq:cgsRG}
c(r,g)= \hat c(\bar g) \exp\left \{-\int_1^r d\log r'\, \gamma_{\phi^4}(\bar g(r',g))\right \}\,.
\eeq
The function $\hat c(\bar g)= c(1,\bar g)$ here is supposed to be obtained by matching with perturbation theory, i.e.~by setting $g\to \bar g$, $r\to 1$ in \reef{eq:cgs}. So we have
\beq
\label{eq:hatc}
\hat c(g) = Y(-\eps g+ (3/2)K g^2)+O(g^3)\,.
\eeq
Recall that $g_*\approx \eps/K$, and thus $c(g_*)>0$. We see thus that $c(r,g)$ starts out negative in the UV and then becomes positive in the IR. So there is a chance that $C(x)$ integrates to zero. Let's see this in detail.

We can perform a couple more steps in full generality.
First of all notice that the argument of the exponential in \reef{eq:cgsRG} can be also written as
\beq
\exp\left \{\int_g^{\bar g} dt\,\frac{\gamma_{\phi^4}(t)}{\beta(t)}\right\}\,.
\eeq
Now, we have to all orders in perturbation theory (see \cite{Paulos:2015jfa}, (6.9); notice that this is consistent with the lowest order expression given in \reef{eq:Zphi4})
\beq
Z_{\phi^4}=-\beta(g) \mu^\eps/(\eps g_0)\,,
\eeq
from where it follows
\beq
\gamma_{\phi^4}=\eps +\beta'(g)\,.
\eeq
This allows to perform exactly the integral in $t$ in \reef{eq:cgsRG}. We thus get
\beq
\label{eq:cgs1}
c(g,s)= \hat c(\bar g) s^{-\eps} \beta(\bar g)/\beta(g)\,.
\eeq
Using this expression the integral of the OPE kernel 
\beq
I = \int d^dx\, C(x,\mu,g) = \Sd \int_0^\infty \frac{dx}{x^{1-\eps}} c(\mu x,g)\,
\eeq
can be dramatically simplified. Indeed, plugging in \reef{eq:cgs1} we have
\beq
I = \frac{ \Sd \mu^\eps}{\beta(g)} I_0,\quad I_0 = \int_0^\infty d\log r\, \hat c(\bar g(r,g)) \beta(\bar g) = \int _0^{g_*} dt\,\hat c(t)
\eeq
Plugging in \reef{eq:hatc}, we get $I_0=0$ to the order we are working. This is a nontrivial check as it involves cancelation of two order $\eps^2$ contributions.

\section{Checks of the OPE ratio relations}
\label{sec:checksOPE}

\subsection{$\phi^4$-flow}
\label{sec:checksOPEphi4}

Eq.~\reef{eq:ratio} predicts the following dependence of the OPE coefficient ratio on $\Delta_{1}$ and $\Delta_2$:
\begin{equation} \label{ratiogphi4}
R_{12}  \equiv \frac{   \lambda_{12 \tilde \phi}} {\lambda_{12 \tilde \phi^3 }   } = g  \cdot  \frac{\Gamma(\frac{\Delta_\phi + \Delta_{12}}{2}) \Gamma(\frac{\Delta_\phi - \Delta_{12}}{2}) }{\Gamma(\frac{\Delta_{\phi^3} + \Delta_{12}}{2}) \Gamma(\frac{\Delta_{\phi^3} - \Delta_{12}}{2})}\cdot A(g) \,.
\end{equation}
The function $A(g)$ has a finite limit for $g \to 0$. It is independent of $\Delta_1$ and $\Delta_2$ (but it does depend on $\Delta_\phi$ and $d$). Here we would like to perform some very simple checks of this relation.

Scalar conformal primaries of the mean field theory schematically have the form:
\beq
\calO\sim \del^{2k}\phi^{n},
\eeq
where derivatives have to be distributed and contracted to get a primary. We take two primaries of this form, assuming $n_1\ge n_2$ without loss of generality. To get a possibility for nonzero 3pt functions we assume that $n_1-n_2$ is odd. 

Eq.~\reef{ratiogphi4} generically predicts $R_{12}=O(g)$. However, there are exceptions if 
\begin{enumerate}
\item $n_1=n_2+1$, $k_1\ge k_2$, 
\item $n_1=n_2+3$, $k_1\ge k_2$.
\end{enumerate} 
In case 1(resp.~2) the second Gamma functions in the numerator, resp.~denominator, is near a pole. Assuming that $\calO_1$ and $\calO_2$ get unequal anomalous dimensions $O(g)$ we expect $\Gamma\sim 1/g$. This predicts 
\beq
\label{eq:predict}
\text{case 1: } R_{12}=O(1), \qquad \text{case 2: } R_{12}=O(g^2)\,.
\eeq
We would like to check, in several simple examples, how this agrees with the perturbation theory.

First consider $k_1=k_2=0$. Let's start with $\calO_1=\phi^{n+1}$, $\calO_2=\phi^n$ (case 1). In perturbation theory, 3pt functions $\vev{\phi \phi^{n+1}\phi^n}$ and $\vev{\phi^3 \phi^{n+1}\phi^n}$ exist already in the mean field theory. Here are the corresponding diagrams for $n=4$: 
\beq
 \includegraphics[width=.2\textwidth]{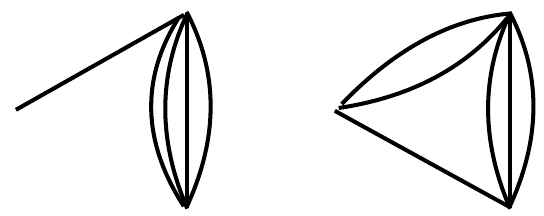}
\eeq
Thus $R_{12}=O(1)$, in agreement with the above prediction.

Next let us examine $\calO_1=\phi^{n+3}$, $\calO_2=\phi^n$ (case 2). In perturbation theory, 3pt functions $\vev{\phi^3 \phi^{n+3}\phi^n}$ appears at $O(1)$, while $\vev{\phi \phi^{n+3}\phi^n}$ needs one coupling insertion. E.g.~for $n=2$: 
\beq
 \includegraphics[width=.2\textwidth]{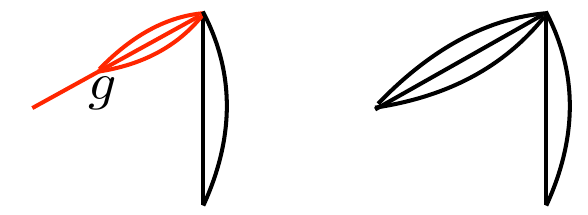}
\eeq
Moreover, the integral over the position of this insertion (the red subdiagram) gives an extra $O(\eps)$ suppression, thanks to the formula:
\begin{equation}
\label{eq:key}
\int d^d x  \, \frac{1}{|x + y|^{2a}  |x|^{2b}    } \sim \frac{1}{\Gamma (d - a - b)} \, ,
\end{equation}
where in the considered case $a + b= d+O(\eps)$.\footnote{The physical reason for this suppression is that the integral becomes conformal for $\eps=0$, and can be interpreted as the leading correction to the 2pt function $\vev{\phi \phi^3}$, so it must vanish for $\eps=0$ since conformal invariance forbids nonzero 2pt functions of operators of unequal dimension.} So all in all we have $R_{12}=O(\eps g)=O(g^2)$, in  agreement with \reef{eq:predict}.

Now let's consider $\calO_2=\phi^{n}$, $\calO_1=\phi^{n+5+2r}$. In perturbation theory, both 3pt functions require coupling insertions: 
\beq
\vev{\phi \calO_1\calO_2}_g=O(g^{r+2}),\qquad \vev{\phi^3 \calO_1\calO_2}_g=O(g^{r+1})\,.
\eeq
Both have just one power of $\eps$ suppression due to ``red subdiagrams''. So $R_{12}=O(g)$ as expected for the generic case. Here's an example for $r=0$:
\beq
 \includegraphics[width=.2\textwidth]{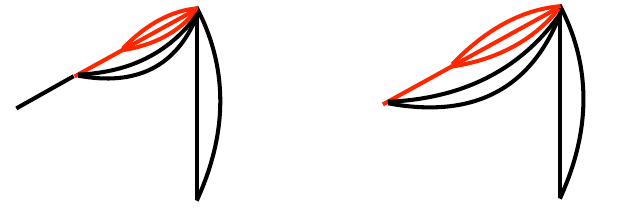}
\eeq

Let us now examine a more complicated example. We would like to understand the origin of the restriction $k_1\ge k_2$. We will only study $n_1=n_2+1$ but $k_1,k_2$ general. The 3pt function $\vev{\phi \calO_1\calO_2}$ is nonzero in the mean field theory if and only if $\calO_1$ occurs in the OPE $\phi(x)\times \calO_2(0)$. To pick up terms with $n_1=n_2+1$ $\phi$'s in the OPE, we are not allowed to take any Wick contractions, only to expand $\phi(x)$ around $x=0$. This will produce terms with a non-negative number of derivatives acting on $\phi$, i.e.~all $\calO_1$'s of this type have $k_1\ge k_2$. Analogously it's easy to see that $\calO_1\in \phi^3\times \calO_2$ is only possible if $k_1\ge k_2$. 

So we get the following picture. If $k_1\ge k_2$, then both 3pt functions $\vev{\phi \calO_1\calO_2}$, $\vev{\phi^3 \calO_1\calO_2}$
exist in MFT, we expect $R_{12}=O(1)$, and this is what \reef{eq:predict} predicts in this case.
On the other hand if $k_1 < k_2$, then we need coupling insertions to generate the 3pt functions. Generically we expect $\vev{\phi^3 \calO_1\calO_2}_g=O(g)$. On the other hand, 
$\vev{\phi \calO_1\calO_2}_g=O(g^2)$, since the diagrams with just one coupling insertions will vanish, as they involve $\vev{\phi^3 \calO_1\calO_2}_{\rm MFT}=0$. So we indeed expect $R_{12}=O(g)$, consistently with \reef{eq:predict}.

\subsection{$\sigma\chi$-flow}
\label{sec:checksLeonardo}
  
 This section is the analogue of the previous one for the $\sigma\chi$-flow at $s\to s_*$. The OPE coefficient ratio is predicted by (\ref{dualratio}) to have dependence of the ratio on $\Delta_{1}$ and $\Delta_2$ of the form:
\begin{equation} \label{ratiog}
R_{12}  \equiv \frac{   \lambda_{12 \tilde \chi }} {\lambda_{12 \tilde \sigma }   } = g  \cdot  \frac{\Gamma(\frac{\Delta_\chi + \Delta_{12}}{2}) \Gamma(\frac{\Delta_\chi - \Delta_{12}}{2}) }{\Gamma(\frac{\Delta_\sigma + \Delta_{12}}{2}) \Gamma(\frac{\Delta_\sigma - \Delta_{12}}{2})}\cdot \tilde A(g) \,.
\end{equation}
To avoid awkward square roots we have expressed the answer as a 
 function of the  coupling $g$, to be evaluated at the fixed point $g_* \sim \sqrt{\delta}$.
The function $\tilde A(g)$ has a finite limit for $g \to 0$. It is independent of $\Delta_1$ and $\Delta_2$ (but it does depend on $\Delta_\sigma$, $\Delta_\chi=d-\Delta_\sigma$ and $d$). In particular, since we will be focussing on the dependence on $\Delta_{1}$ and $\Delta_2$, we will not have to keep track of the IR normalization of $\sigma$ and $\chi$ (as long as it's the same for all considered correlators).

For generic ${\cal O}_1$
and ${\cal O}_2$, the $\chi$ OPE coefficient is suppressed by a power of $g$, reflecting 
the  factorization of the theory  into the product of the SRFP and of the Gaussian $\chi$ model as $s \to s_*$. As we will see, there are however interesting cases where  one of the Gamma functions develops a pole, corresponding to a non-vanishing $\chi$ correlator in the factorized
theory (if the pole is in one of the numerator Gamma functions) or to a further suppression of the  $\chi$ correlator in conformal perturbation theory  (if the pole is in one of the denominator Gamma functions).

 Let $ \{ {\cal O}^{\rm SR}_{\Delta_i, \ell_i, \alpha_i}  \}$ be the conformal primary operators of the SRFP, where $\Delta_i$ and $\ell_i$ are their
 conformal dimensions and spins, and $\alpha_i = \pm 1$ their $\mathbb{Z}_2$ quantum numbers. The general scalar  primary of the factorized theory at $s = s_*$ takes the schematic form
 \begin{equation} \label{generic}
\partial^{ \ell + 2k}[ {\cal O}^{\rm SR}_{\Delta_, \ell, \alpha}    \chi^n] \, ,
\end{equation}
where the derivatives are contracted to give a scalar and distributed in such a way to give a primary. The $\chi$ parity $\beta = (-1)^n$ is  an exact $\mathbb{Z}_2$ symmetry in the factorized theory. The interaction $g \int \chi \sigma$ preserves the diagonal $\mathbb{Z}_2$ symmetry, whose quantum number we denote by $\nu = \alpha \cdot \beta$.  As we turn on $g$, two operators with different ${\cal O}^{\rm SR}$ and different numbers of $\chi$'s can a priori mix,
provided that they have the same conformal dimension
and that $\alpha_1 \beta_1 = \alpha_2 \beta_2$.  In such a case, the correct dilation eigenstates will be linear combinations of states of the form (\ref{generic}). 
This is however a very rare and perhaps impossible phenomenon, which will discuss  at the end. In the bulk of the analysis, we will assume that both ${\cal O}_1$ and ${\cal O}_2$ take the form (\ref{generic}),\footnote{A common phenomenon is the mixing of states of the form  (\ref{generic}) with the same  ${\cal O}^{\rm SR}$ and the same $n$, $k$ and $\ell$, but different ways to distribute the derivatives.}
\begin{equation}
{\cal O}_1 = \partial^{ \ell_1 + 2k_1}   [{\cal O}^{\rm SR}_{\Delta_1, \ell_1, \alpha_1} \chi^{n_1}] \, , \quad {\cal O}_2 = \partial^{ \ell_2 + 2k_2} [{\cal O}^{\rm SR}_{\Delta_2, \ell_2, \alpha_2}   \chi^{n_2}] \, .
\end{equation}
In order for the 3pt functions $\langle {\cal O}_1 {\cal O}_2  \sigma \rangle$  and $\langle {\cal O}_1 {\cal O}_2  \chi \rangle$ to have a chance of being non-zero, the diagonal $\mathbb{Z}_2$'s of ${\cal O}_1$ and  ${\cal O}_2$ must be opposite, $\nu_1 = - \nu_2$, and we will assume that this is always the case.

 In the factorized theory at $s = s_*$,  $\sigma$ and $\chi$ have different $\mathbb{Z}_2 \times \mathbb{Z}_2$ quantum numbers, respectively $(- , +)$ and $(+, -)$,  so if $\langle {\cal O}_1 {\cal O}_2  \sigma \rangle$ is non-zero,
$\langle {\cal O}_1 {\cal O}_2  \chi \rangle$ is zero, and viceversa. We then expect the ratio (\ref{ratiog}) to be either zero or infinite as $g\to0$. 

With no loss of generality, we will assume that $n_1 \geq n_2$. Let's enumerate the various cases:
\begin{enumerate}
\item  $n_1 = n_2$.

 In this case, the scaling in $g$ as $g \to 0$ expected from $\mathbb{Z}_2 \times \mathbb{Z}_2$ selection rules is given, generically,\footnote{One exceptional case is $\calO_1=\calO_1^{\rm SR}\calO_1^\chi$ and  $\calO_2=\calO_2^{\rm SR}\calO_2^\chi$ with $\calO_1^\chi$ and $\calO_2^\chi$ two different $\chi$ theory primaries which contain the same number of $\chi$'s. In this case $\langle {\cal O}_1 {\cal O}_2  \sigma\rangle$ vanishes in the factorized theory, and the analysis needs to be modified.} by
\begin{equation}
\langle {\cal O}_1 {\cal O}_2  \sigma \rangle_g = O(1) \, , \quad \langle {\cal O}_1 {\cal O}_2  \chi \rangle_g  = O(g) \, .
\end{equation}
 Indeed $\alpha_1 = - \alpha_2$, making 
$\langle {\cal O}^{\rm SR}_1 {\cal O}^{\rm SR}_2 \sigma \rangle \neq 0$ in the SRFP.  While $\langle {\cal O}_1 {\cal O}_2  \sigma \rangle$ is already non-zero at $g=0$,
a non-zero  $\langle {\cal O}_1 {\cal O}_2  \chi \rangle$ requires
 one insertion of the interaction $g \int \chi \sigma$. We have $\Delta_{12} = \Delta^{\rm SR}_1 - \Delta^{\rm SR}_2 + 2k_1 - 2k_2$ at $g=0$.  In the generic case, there are no poles coming from 
  the Gamma functions of (\ref{ratiog}),  and the $O(g)$ behavior of the ratio expected from conformal perturbation theory is correctly reproduced.   We have checked in simple examples that the numerical coefficients are also correctly reproduced.
    
No two operators in the SRFP
 have dimensions differing by $\Delta_\chi$ plus an integer, so for $n_1 = n_2$ there are never poles from the numerator Gamma functions. It is possible however
 to have a pole from the denominator Gamma functions, resulting in a zero in $R_{12}$, if $\Delta_{12} = \Delta_\sigma + 2k_1 - 2 k_2$ with $k_1 \geq k_2$,
 or  $\Delta_{12} = -\Delta_\sigma + 2k_1 - 2 k_2$ with $k_1 \leq k_2$.  Given that operators with given $\mathbb{Z}_2 \times \mathbb{Z}_2$ quantum numbers
 acquire an $O(g^2)$ anomalous dimension, this zero should be interpreted as an additional factor of $g^2$, so all in all $R_{12} = O(g^3)$ in these cases.
 Let's see in an example how this can be compatible with conformal perturbation theory. Take
 \begin{equation} \label{firstdimreg}
 {\cal O}_1 = \sigma  \chi^n  \, ,\quad   {\cal O}_2 =  \chi^n \, .
 \end{equation}
We still have $\langle {\cal O}_1 {\cal O}_2  \sigma \rangle_g = O(1)$, 
but we are going to argue
that $\langle {\cal O}_1 {\cal O}_2  \chi \rangle_g = O(g^3)$. 
Indeed, the $O(g)$ 
contribution to $\langle {\cal O}_1 {\cal O}_2  \chi \rangle_g$ arises from the integral
\begin{equation} \label{renormalizediszero}
g \int d^d y  \; \langle   \sigma \chi^n (x_1)  \;   \chi^n(x_2) \;  \chi(x_3) \; \sigma  \chi (y) \rangle_0 \, ,
\end{equation}
which needs to be regulated and renormalized. We claim that the renormalized integral is actually $O(\delta)$, so that the net contribution is $O(g \delta) = O(g^3)$. A quick and dirty way to see this is to evaluate the integral in dimensional regularization, using Eq.~\reef{eq:key}. Proving this for examples more complicated than (\ref{firstdimreg}) requires a version of \reef{eq:key} with an extra $(x+z)^{2c}$ in the numerator.

  \item  $n_1 = n_2+1$.  
  
Since we require $\nu_1 = - \nu_2$, in this case we have $\alpha_1 = \alpha_2$. In the generic case,   ${\cal O}^{\rm SR}_1$ and ${\cal O}^{\rm SR}_2$ are two {\it different} operator with the same SR $\mathbb{Z}_2$ quantum number, and then 
\begin{equation}
\langle {\cal O}_1 {\cal O}_2  \sigma \rangle_g = O(g) \, , \quad \langle {\cal O}_1 {\cal O}_2  \chi \rangle_g  = O(g^2) \, .
\end{equation}
Indeed, for the first correlator one needs a single insertion of the interaction, while the second correlator vanishes at $g=0$ (because  ${\cal O}^{\rm SR}_1 \neq {\cal O}^{\rm SR}_2$) and then selection rules force the insertion of {\it two} interactions. We observe again the generic behavior $O(g)$ for the ratio of OPE coefficients, in agreement with (\ref{ratiog}).

The more interesting case is when  ${\cal O}^{\rm SR}_1 = {\cal O}^{\rm SR}_2$. Then selection rules would predict
\begin{equation} \label{funnycasegeneric}
\langle {\cal O}_1 {\cal O}_2  \sigma \rangle_g = O(g) \, , \quad \langle {\cal O}_1 {\cal O}_2  \chi \rangle_g  = O(1) \, .
\end{equation}
In this case, $\Delta_{12} = \Delta_\chi + 2k_1 - 2k_2$ for $g=0$. If $k_1 \geq k_2$, we encounter in a pole from the second Gamma function in the numerator of (\ref{ratiog}).
Operators with definite $\mathbb{Z}_2 \times \mathbb{Z}_2$ quantum numbers  acquire anomalous dimension at order $O(g^2)$, so for small $g$ the pole is regulated to $\sim 1/g^2$, and $R_{12} \sim 1/g$, in  agreement with (\ref{funnycasegeneric}). On the other hand, if $k_1 < k_2$, we seem to have a problem. There is {\it no} pole in the Gamma function, so (\ref{ratiog}) predicts $R_{12} \sim O(g)$, in contradiction with (\ref{funnycasegeneric}).

To understand the resolution of this puzzle, let's consider the following example. 
Let
\begin{equation}
{\cal O}_1 = \sigma \chi^3 \, , \quad {\cal O}^A_2 = \sigma (\partial_\mu \chi \partial^\mu \chi + a \chi \Box \chi ) \, , \quad {\cal O}^B_2 = \partial_\mu \sigma  \partial_\mu \chi \chi + b \Box \sigma  \chi^2 \, ,
\end{equation}
where the coefficients $a$ and $b$ are fixed such that  ${\cal O}^A_2$ and ${\cal O}^B_2$ are conformal primaries. 
It is important to realize that ${\cal O}^A_2$ and ${\cal O}^B_2$ mix in conformal perturbation theory, indeed the $O(g^2)$ correction to the dilation operator contains an off-diagonal term arising from the non-vanishing correlator
$ \langle {\cal O}^A_2  \, {\cal O}^B_2 \, \sigma \chi   \sigma \chi  \rangle_0$.
The eigenstates of the dilation operator take the form
\begin{equation} \label{dilationeigenstates}
 {\cal O}^I_2   =    {\cal O}^A_2    +  c  {\cal O}^B_2         \, \quad  {\cal O}^{II}_2 =  {\cal O}^A_2    +  d  {\cal O}^B_2     \,,
\end{equation}
 where the coefficients $c$ and $d$ have no $g$ dependence.
We claim that 
\begin{equation} 
\langle {\cal O}_1  {\cal O}_2  \sigma \rangle  = O(g) \, ,  \quad  \langle {\cal O}_1  {\cal O}_2  \chi \rangle  = O(g^2) \, ,
\end{equation} 
where ${\cal O}_2$ is either one of the dilation eigenstates in (\ref{dilationeigenstates}), so  that $R_{12} = O(g)$ in agreement with (\ref{ratiog}).
Indeed 
\begin{equation} \label{samereason}
\langle  {\cal O}_1 (x_1) {\cal O}_2  (x_2) \chi (x_3) \rangle_0 \sim \langle  \chi(x_1) \chi(x_3) \rangle_0 \,   \langle  \sigma \chi^2(x_1) {\cal O}_2 (x_2) \rangle_0= 0 \, ,
\end{equation}
where in the last step we have used orthogonality of conformal primaries of different dimension.
We then need to go to $O(g^2)$ to find a non-zero contribution to  $\langle {\cal O}_1 {\cal O}_2  \chi \rangle_g$. On the other hand,
\beq
\label{itsnotzero}
\langle  {\cal O}_1 (x_1) {\cal O}_2  (x_2) \sigma  (x_3)  \sigma \chi (x_4) \rangle_0  \sim    \langle  \chi(x_1) \chi(x_4) \rangle_0 \,   \langle  \sigma \chi^2(x_1) {\cal O}_2 (x_2)  \sigma  (x_3) \sigma (x_4) \rangle_0
\neq 0 \, ,
\eeq
so  $\langle {\cal O}_1 {\cal O}_2  \sigma \rangle_g = O(g)$. It is essential that the dilation eigenstate ${\cal O}_2 \neq {\cal O}^A_2$, {\it i.e.} that it does {\it not} factorize into $\sigma$ times a primary of the $\chi$ theory, otherwise the correlator (\ref{itsnotzero}) would vanish.

  \item  $n_1 = n_2+ 2 m$, $m \geq 1$.
  We must have $\alpha_1 = -\alpha_2$. Selection rules and the requirement that we have enough $\chi$ insertions 
would naively give
 \begin{equation}
 \langle {\cal O}_1 {\cal O}_2  \sigma \rangle_g = O(g^{2m}) \, , \quad \langle {\cal O}_1 {\cal O}_2  \chi \rangle_g  = O(g^{2m-1}) \, ,
 \end{equation}
  but the actual behavior is
   \begin{equation}
 \langle {\cal O}_1 {\cal O}_2  \sigma \rangle_g = O(g^{2m}) \, , \quad \langle {\cal O}_1 {\cal O}_2  \chi \rangle_g  = O(g^{2m+1}) \, .
 \end{equation}
 One can convince oneself in examples that the $O(g^{2m-1})$ contribution to  $\langle {\cal O}_1 {\cal O}_2  \chi \rangle_g$ vanishes -- the corresponding renormalized
 correlator turns out to be $O(\delta)$, a phenomenon we already encountered in (\ref{renormalizediszero}). It would be nice to find a general argument. 
 Conformal perturbation theory is then consistent with the generic
 behavior $R_{12} = O(g)$ predicted by (\ref{ratiog}).
   
    \item  $n_1 = n_2+ 2 m +1$, $m \geq 1$.
    Now $\alpha_1 = \alpha_2$. The naive scaling is
     \begin{equation}
 \langle {\cal O}_1 {\cal O}_2  \sigma \rangle_g = O(g^{2m+1}) \, , \quad \langle {\cal O}_1 {\cal O}_2  \chi \rangle_g  = O(g^{2m}) \, ,
 \end{equation}
 and the correct one
   \begin{equation}
 \langle {\cal O}_1 {\cal O}_2  \sigma \rangle_g = O(g^{2m+1}) \, , \quad \langle {\cal O}_1 {\cal O}_2  \chi \rangle_g  = O(g^{2m+2}) \, ,
 \end{equation}
 for the same reason as the previous case.    Again, we find agreement with (\ref{ratiog}).

\end{enumerate}

 Finally, let's consider the possibility of mixing of states of the form (\ref{generic}) with different $\mathbb{Z}_2 \times \mathbb{Z}_2$ quantum numbers. 
 This requires a conspiracy of quantum numbers that is hard to arrange. A naive candidate is the following.
 Take $d=2$
 and consider
 \begin{equation}
 {\cal O}_A =  (\partial_\mu \chi  \partial_\mu \chi + a\, \chi \Box \chi) \chi^{n-2}  \, ,\quad  {\cal O}_B = \sigma \chi^{n+1}\, ,
 \end{equation}
where the coefficient $a$ is fixed to make  ${\cal O}_A$ a conformal primary. Since $\Delta_\chi + \Delta_\sigma = 2$, 
both states have dimension $2 + n \Delta_\chi$. At first sight, it might seem that  dilation operator has an off-diagonal component at order $O(g)$ arising
from 
\begin{equation} \label{AB}
g \int d^2 y \, \langle  {\cal O}_A  (x_1) {\cal O}_B  (x_2)\, \chi \sigma (y) \rangle\, .
\end{equation}
However, this correlator is of order $O(\delta)$ for the same reason as the correlator in (\ref{samereason}). An off-diagonal component would arise
at $O(g^3)$, but there are diagonal anomalous dimensions already at order $O(g^2)$, which presumably lift the degeneracy between ${\cal O}_A$ and ${\cal O}_B$. So it appears that there is no mixing after all,
barring some coincidence. It would be nice to decide this one way or another by a detailed computation.

Let's consider the power counting in the scenario that the two states  (\ref{AB}) {\it do} mix to leading order. (Maybe a more intricate example would actually work along these lines).
Let ${\cal O}_1$ be one of the two dilation eigenstates, of the form
\begin{equation}
{\cal O}_1 = {\cal O}_A + b\,  {\cal O}_B 
\end{equation}
for some coefficient $b = O(1)$,
and choose
\begin{equation}
{\cal O}_2 = \chi^n \sigma \,.
\end{equation}
Then we have
\begin{equation}
\langle {\cal O}_1 {\cal O}_2  \sigma \rangle_g = O(1) \, , \quad \langle {\cal O}_1 {\cal O}_2  \chi \rangle_g  = O(1) \, ,
\end{equation}
so $R_{12}$ would be finite as $g \to 0$ limit in this  scenario. Note that $\Delta_{12} = \Delta_\chi$ for $g=0$, which gives rise to a pole in (\ref{ratiog}). Now however the anomalous dimension of ${\cal O}_1$ is of order $O(g)$. The pole is regulated to $\sim 1/g$ and
(\ref{ratiog}) predicts $R_{12} = O(1)$. So  we win again.

\bibliography{LRI-biblio.bib}
\bibliographystyle{utphys.bst}
\end{document}